\documentclass{article}
\usepackage[psamsfonts]{amsfonts}
\usepackage{amsfonts,amssymb,amsmath,stmaryrd}
\usepackage{theorem}
\usepackage{xspace}
\usepackage{epsfig}
\usepackage{subfigure}

\usepackage[bookmarksnumbered=true,%
            pdftitle={Partitioning the Threads of a Mobile System},
            pdfauthor={J{\'e}r{\^o}me Feret},%
            pdfkeywords={static analysis, 
                         mobile systems,
                         abstract interpretation}]{hyperref}

\newcommand{\bydef}{:=}

\newcommand{\fracbis}[2]{\frac{\strut #1}{\strut #2}}
\newcommand{\boxexample}{$\Box$}

\newtheorem{thm}{Theorem}[section]
\newtheorem{example}[thm]{Example}
\newtheorem{definition}[thm]{Definition}
\newtheorem{theorem}[thm]{Theorem}

\newcommand{\picalcul}{$\pi$-calculus}

\newcommand{\ambient}{\emph{ambient}}
\newcommand{\ambients}{\emph{ambients}}
\newcommand{\Ambients}{\emph{Ambients}}
\newcommand{\spicalcul}{{\emph{spi}-calculus}}
\newcommand{\bioambients}{{\textsc{bio}-\emph{ambients}}}

\newcommand{\Names}{\mathcal{V}}
\newcommand{\Labels}{\mathcal{L}}
\newcommand{\concu}{|}
\newcommand{\nuu}{\nu\;}
\newcommand{\repli}{\ast}
\newcommand{\rec}{?}
\newcommand{\eme}{!}
\newcommand{\emptypro}{\mathbf{0}}

\newcommand{\pp}[1]{$\bf{#1}$}
\newcommand{\pps}[1]{(at program point \pp{#1})}
\newcommand{\make}{\mathrm{make}}
\newcommand{\bang}[1]{\overline{\repli}^{#1}}
\newcommand{\globalname}[1]{\mathrm{#1}}
\newcommand{\internal}[1]{\textnormal{\texttt{#1}}}
\newcommand{\capability}[1]{\textsc{#1}}
\newcommand{\data}[1]{\textsf{#1}\,}
\newcommand{\recur}[1]{{\textsl{#1}}}
\newcommand{\variable}[1]{\textit{#1}}

\newcommand{\cack}{\variable{ack}}
\newcommand{\cdata}{\data{data}}
\newcommand{\addinit}{\data{add}}
\newcommand{\addread}{\data{return}}
\newcommand{\addwrite}{\data{ack}}
\newcommand{\port}{\variable{fwd}}
\newcommand{\cbang}{\recur{rec}}
\newcommand{\ccreate}{\globalname{alloc}}
\newcommand{\cnull}{\globalname{null}}
\newcommand{\ccell}{\internal{cell}}
\newcommand{\cread}{\capability{read}}
\newcommand{\cwrite}{\capability{write}}
\newcommand{\datawrite}{\variable{val}'}
\newcommand{\dataread}{\variable{val}}

\newcommand{\memlaba}{1}
\newcommand{\memlabb}{2}
\newcommand{\memlabc}{3}
\newcommand{\memlabd}{4}
\newcommand{\memlabe}{5}
\newcommand{\memlabf}{6}
\newcommand{\memlabg}{7}
\newcommand{\memlabh}{8}
\newcommand{\memlabi}{9}
\newcommand{\memlabj}{10}
\newcommand{\memlabk}{11}
\newcommand{\memlabl}{12} 
\newcommand{\memlabm}{13} 
\newcommand{\memlabn}{14} 
\newcommand{\memlabo}{15}  
\newcommand{\memlabp}{16}  
\newcommand{\memlabq}{17} 
\newcommand{\memlabr}{18}  
\newcommand{\memlabs}{19}  
\newcommand{\memlabt}{20}

\newcommand{\crec}{\recur{rec}}

\newcommand{\cfalaba}{1}
\newcommand{\cfalabc}{2}
\newcommand{\cfalabd}{3}
\newcommand{\cfalabe}{4}
\newcommand{\cfalabf}{5}
\newcommand{\cfalabg}{6}
\newcommand{\cfalabh}{7}
\newcommand{\cfalabi}{8}
\newcommand{\cfalabj}{9}
\newcommand{\cfalabk}{10}
\newcommand{\cfalabl}{11}
\newcommand{\cfalabm}{12}
\newcommand{\cfalabn}{13}
\newcommand{\cfalabo}{14}
\newcommand{\cfalabp}{15}
\newcommand{\llinkint}[1]{\internal{l}{#1}}
\newcommand{\rlinkint}[1]{\internal{r}{#1}}
\newcommand{\clinkint}[1]{\internal{c}{#1}}

\newcommand{\celllinkint}[1]{[\llinkint{#1},\clinkint{#1},\rlinkint{#1}]}

\newcommand{\llinkvar}[1]{\internal{l}{#1}}
\newcommand{\rlinkvar}[1]{\internal{r}{#1}}
\newcommand{\clinkvar}[1]{\internal{c}{#1}}

\newcommand{\cset}{\capability{set}}
\newcommand{\celllinkvar}[1]{[\llinkvar{#1},\clinkvar{#1},\rlinkvar{#1}]}

\newcommand{\flowdom}[1]{\mathcal{F}(#1)}
\newcommand{\flowsub}[1]{\sqsubseteq_{\flowdom{#1}}}
\newcommand{\flowconc}[1]{\gamma_{\flowdom{#1}}}
\newcommand{\flowbot}[1]{\bot_{\flowdom{#1}}}
\newcommand{\normalflowdom}[1]{\mathcal{F}_{n}(#1)}
\newcommand{\normalflowsub}[1]{\sqsubseteq_{\normalflowdom{#1}}}
\newcommand{\normalflowconc}[1]{\gamma_{\normalflowdom{#1}}}
\newcommand{\normalflowcup}[1]{\sqcup_{\normalflowdom{#1}}}
\newcommand{\cana}{\internal{a}}
\newcommand{\semlaba}{1}
\newcommand{\semlabb}{2}
\newcommand{\semlabc}{3}
\newcommand{\semlabd}{4}
\newcommand{\semlabe}{5}

\newcommand{\alloc}{\globalname{b}}
\newcommand{\addone}{\variable{c}}
\newcommand{\addtwo}{\variable{c}'}
\newcommand{\addthree}{\variable{c}''}
\newcommand{\argone}{\internal{c}}
\newcommand{\argtwo}{\internal{c}'}
\newcommand{\argthree}{\internal{c}''}
\newcommand{\clock}{\internal{l}}
\newcommand{\cm}{\internal{m}}
\newcommand{\crr}{\internal{r}}
\newcommand{\twolaba}{1}
\newcommand{\twolabb}{2}
\newcommand{\twolabc}{3}
\newcommand{\twolabd}{4}
\newcommand{\twolabe}{5}
\newcommand{\twolabf}{6}
\newcommand{\twolabg}{7}
\newcommand{\twolabh}{8}
\newcommand{\twolabi}{9}
\newcommand{\twolabj}{10}
\newcommand{\twolabk}{11}
\newcommand{\twolabl}{12}
\newcommand{\twolabm}{13}
\newcommand{\twolabn}{14}
\newcommand{\twolabo}{15}
\newcommand{\twolabp}{16}
\newcommand{\twolabq}{17}
\newcommand{\twolabr}{18}

\newcommand{\canb}{\internal{b}}
\newcommand{\canc}{\internal{c}}
\newcommand{\varu}{\variable{u}}
\newcommand{\varv}{\variable{v}}
\newcommand{\synlaba}{1}
\newcommand{\synlabb}{2}
\newcommand{\synlabc}{3}
\newcommand{\synlabd}{4}
\newcommand{\synlabe}{5}
\newcommand{\synlabf}{6}
\newcommand{\synlabg}{7}

\newcommand{\server}{\globalname{server}}
\newcommand{\duplicate}{\globalname{give\_id}}
\newcommand{\instance}{\globalname{instance}}
\newcommand{\client}{\globalname{client}}
\newcommand{\ambk}{\internal{k}}
\newcommand{\ambx}{\internal{x}}
\newcommand{\ambp}{\internal{p}}
\newcommand{\amblaba}{1}
\newcommand{\amblabb}{2}
\newcommand{\amblabc}{3}
\newcommand{\amblabd}{4}
\newcommand{\amblabe}{5}
\newcommand{\amblabf}{6}
\newcommand{\amblabg}{7} 
\newcommand{\amblabh}{8} 
\newcommand{\amblabi}{9} 
\newcommand{\amblabj}{10} 
\newcommand{\amblabk}{11} 
\newcommand{\amblabl}{12} 
\newcommand{\amblabm}{13} 
\newcommand{\amblabn}{14} 
\newcommand{\amblabo}{15} 
\newcommand{\amblabp}{16} 
\newcommand{\amblabq}{17} 
\newcommand{\amblabr}{18} 
\newcommand{\bfclient}{\globalname{\bf client}}
\newcommand{\bfduplicate}{\globalname{\bf giveid}}
\newcommand{\bfp}{\internal{\bf p}}
\newcommand{\bfserver}{\globalname{\bf server}}

\newcommand{\emptyfun}{\emptyset}
\newcommand{\prolabel}{\Labels}
\newcommand{\System}{\mathcal{S}}

\newcommand{\FN}{\texttt{fv}}
\newcommand{\interface}{\mathtt{I}}
\newcommand{\type}{\texttt{type}}
\newcommand{\Extract}{\mathbf{\beta}}

\newcommand{\Markers}{\Labels^{\ast}}
\newcommand{\Env}[1]{\Markers \times (#1 \rightarrow (\Names\times\Markers))}
\newcommand{\env}[1]{\mathcal{E}(#1)}
\newcommand{\initstan}{\mathcal{I}}
\newcommand{\nonstan}[1]{\overset{\scriptscriptstyle{#1}}{\scriptstyle \longrightarrow}}
\newcommand{\states}{\mathcal{C}}
\newcommand{\Coll}{\states(\System)}
\newcommand{\component}{\texttt{comp}}
\newcommand{\inputm}{\emph{input}}
\newcommand{\outputm}{\emph{output}}
\newcommand{\fetch}{\emph{fetch}}
\newcommand{\args}{\texttt{arg}}
\newcommand{\canal}{\texttt{chan}}
\newcommand{\cont}{\texttt{cont}}
\newcommand{\launch}{\texttt{launch}}
\newcommand{\lrec}{l_{\rec}}
\newcommand{\leme}{l_{\eme}}
\newcommand{\idrec}{\textit{id}_{\rec}}
\newcommand{\ideme}{\textit{id}_{\eme}}
\newcommand{\Erec}{E_{\rec}}
\newcommand{\Eeme}{E_{\eme}}
\newcommand{\trec}{t_{\rec}}
\newcommand{\ttuplerec}{(\lrec,\idrec,\Erec)}
\newcommand{\teme}{t_{\eme}}
\newcommand{\ttupleeme}{(\leme,\ideme,\Eeme)}
\newcommand{\Vrec}{V_{\rec}}
\newcommand{\Veme}{V_{\eme}}
\newcommand{\translabel}{\Labels^2}
\newcommand{\extendedSigma}{\Sigma}
\newcommand{\ExtendedSigma}{\extendedSigma}
\newcommand{\continued}{(cont.)}

\newcommand{\cnonunu}{\shareanalysis\ \continued}

\newcommand{\piunit}{\alpha_\computationunit}
\newcommand{\computationunit}{\textsc{unit}}
\newcommand{\abstractunit}{\textsc{unit}^{\sharp}}
\newcommand{\giveindex}{\texttt{getvar}}

\newcommand{\abstraction}[1]{\mathcal{A}_{#1}}
\newcommand{\abst}[1]{\textsc{post}_{#1}}
\newcommand{\aunion}[1]{\sqcup_{#1}}
\newcommand{\adom}[1]{\mathcal{C}_{#1}^{\sharp}}
\newcommand{\abot}[1]{\bot_{#1}}
\newcommand{\conc}[1]{\gamma_{#1}}
\newcommand{\ainit}[1]{\mathcal{I}_{#1}^{\sharp}}
\newcommand{\wid}[1]{\nabla_{#1}}
\newcommand{\tuple}[1]{(\adom{#1},
                        \aunion{#1},
                        \abot{#1},\conc{#1},
                        \ainit{#1},
                        \abst{#1},
                        \wid{#1})}
\newcommand{\limit}[1]{\llbracket \mathcal{S} 
                       \rrbracket_{\abstraction{#1}}}
 
\newcommand{\counterpart}[1]{\mathbb{F}_{\scriptscriptstyle\!\mathcal{A}_{#1}}^{\sharp}}

\newcommand{\iteraten}[2]{\mathcal{F}_{#2}^{\scriptscriptstyle \wid{#1}}}

\newcommand{\envf}{\textsc{env}}
\newcommand{\Atomekey}{\textit{Atom}}
\newcommand{\Atomeshortkey}{}
\newcommand{\Atome}[1]{\Atomekey(#1)}
\newcommand{\Moleculekey}{\textit{Molecule}}
\newcommand{\Moleculeshortkey}{}
\newcommand{\Molecule}[2]{\Moleculekey(#1,#2)}
\newcommand{\gammaatome}[1]{\gamma^{\Atomeshortkey}_{#1}}
\newcommand{\cupatome}[1]{\sqcup^{\Atomeshortkey}_{#1}}
\newcommand{\botatome}[1]{\bot^{\Atomeshortkey}_{#1}}
\newcommand{\widatome}[1]{\nabla^{\Atomeshortkey}_{#1}}
\newcommand{\gammamolecule}[2]{\gamma^{\Moleculeshortkey}_{({#1},{#2})}}
\newcommand{\botmolecule}[2]{\bot^{\Moleculeshortkey}_{\scriptscriptstyle ({\scriptscriptstyle #1},{\scriptscriptstyle #2})}}
\newcommand{\Envkey}{\mathrm{env}}
\newcommand{\Envdom}{\mathcal{C}^{\sharp}_{\Envkey}}
\newcommand{\gammaenv}{\gamma_{\Envkey}}
\newcommand{\botenv}{\bot_{\Envkey}}
\newcommand{\cupenv}{\sqcup_{\Envkey}}
\newcommand{\sat}{\models}
\newcommand{\jfetchm}{{\textsc{fetch}}}
\newcommand{\fetchm}[2]{\jfetchm(#1,#2)}
\newcommand{\jgc}{\textsc{gc}}
\newcommand{\gc}[2]{\jgc(#1,#2)}
\newcommand{\jdeclare}{{\bf \nu}^{\sharp}}
\newcommand{\declare}{\jdeclare}
\newcommand{\jextension}{\textsc{new}}
\newcommand{\extension}{\jextension}
\newcommand{\jsync}{\textsc{sync}}
\newcommand{\sync}[2]{\jsync(#1,#2)}
\newcommand{\concat}{\bullet}
\newcommand{\initenv}{\ainit{\Envkey}}

\newcommand{\fst}{\textsc{fst}}
\newcommand{\snd}{\textsc{snd}}

\newcommand{\cfadomain}{labels and equalities}
\newcommand{\icfadomain}{\cfadomain}
\newcommand{\ccfadomain}{\cfadomain\ \continued}

\newcommand{\shareanalysis}{the shared memory}

\newcommand{\cflowanalysis}{\shareanalysis\ \continued}

\newcommand{\tp}{\textsc{cu}}
\newcommand{\var}{\mathcal{K}}
\newcommand{\numkey}{\textsc{con}}
\newcommand{\abstnum}{\mathcal{N}(\var)}
\newcommand{\gammanum}{\gamma_{\abstnum}}
\newcommand{\domnum}{\adom{\numkey}}
\newcommand{\cupnum}{\sqcup_{\abstnum}}
\newcommand{\botnum}{\bot_{\abstnum}}
\newcommand{\xinum}{\chi_{\abstnum}}
\newcommand{\initnum}{\ainit{\numkey}}

\newcommand{\addnum}{+^{\sharp}}
\newcommand{\minnum}{-^{\sharp}}
\newcommand{\syncnume}{\textsc{sync}_{\numkey}}

\newcommand{\concnum}{\gamma_{\numkey}}

\newcommand{\widnum}{\wid{\numkey}}
\newcommand{\Numkey}{\text{con}}
\newcommand{\Numdom}{\adom{\Numkey}}
\newcommand{\botNum}{\abot{\Numkey}}
\newcommand{\cupNum}{\aunion{\Numkey}}
\newcommand{\gammaNum}{\gamma_{\Numkey}}

\newcommand{\countdom}{interval and affine constraints}
\newcommand{\icount}{\countdom}
\newcommand{\ccount}{\countdom\ \continued}

\newcommand{\icontentanalysis}{\shareanalysis\ \continued}
\newcommand{\ccontentanalysis}{\shareanalysis\ \continued}

\begin{document}
\bibliographystyle{plain}
\author{J{\'e}r{\^o}me Feret \\
{\'E}cole Normale Sup{\'e}rieure \\
Harvard Medical School\\
jerome\_feret@hms.harvard.edu}

\title{Partitioning the Threads of a Mobile System}

\maketitle

\begin{abstract}
In this paper, we show how thread partitioning helps in proving properties of  mobile systems.
Thread partitioning consists in gathering the threads of a mobile system into several classes. The partitioning criterion is left as a parameter of both the mobility model and the properties we are interested in. Then, we design a polynomial time abstract interpretation-based static analysis that counts the number of threads inside each partition class.
\end{abstract}

\section{Introduction}

A mobile system is a pool of threads that interact with each other.
These interactions dynamically change the system by controlling both the creation and the destruction of links  between threads (by modifying the accesses to channels and/or modifying the spatial configuration).
These interactions also control the creation of threads.
The size of a mobile system may be unbounded.
A mobile system may describe telecommunication networks, reconfigurable systems, \emph{client-server} applications, cryptographic protocols, or biological systems. Several models exist according to the application field and the granularity of the observation level.

We use abstract interpretation \cite{c:thesis,cc:popl77} to derive abstract semantics, which are sound, decidable, but approximate. We use partitioning \cite{cc:popl79,bourdoncle} to separate the  threads according to dynamical information.  The partitioning criterion depends on both the model and the properties of interest. In models based on channeled communications (as in the $\pi$-calculus \cite{milner:polyadic}), we can partition the threads according to the name of the channel they operate on. In models with explicit locations (as in \ambients\ \cite{ambients}), we can partition the threads according to their location in the system. When there are both channeled communications and locations (as in D-$\pi$ \cite{Dpi} or in \bioambients\ \cite{bio}), we partition the threads according to both the channel they operate on and their location.
In more complex cases, the partitioning criterion may be given manually.
For instance, in the \spicalcul, channels are not relevant, so we partition the threads according to the principals that share a session \cite[p:269]{feret:thesis} thanks to some end-user's annotations. Nevertheless, we believe that a better understanding of the problem 
should allow the automatic inference of these annotations.

Our analysis then counts automatically the number of threads inside  each partition class. To get an accurate analysis, we have to relate, for each computation step,  the partition classes of  the threads that interact and  the partition classes of the threads that are created. When analyzing mobile \ambients\ \cite{NiNi00popl}, these relations are given by the model.
This is not the case in less structured models, where a non uniform (i.e.~that distinguishes recursive instances) analysis \cite{feret:sas2000,feret:esop2002,feret:jlap,feret:thesis} of the dynamic linkage between threads is required. To make contents analysis and non uniform analysis collaborate, we locally partition computation steps \cite{mauborgne:rival05} 
according to some assumptions about the partition classes of the threads that interact. Then, we use a coalesced product between both analyses, so that if 
one detects that some assumptions are contradictory, the other ignores the corresponding interaction.

We apply our framework to prove  automatically the absence of race conditions in a shared-memory with dynamic allocation written in the \picalcul. We also analyze precisely the relation between the contents of an \ambient\ and its location in the network. In the author's PhD.~Thesis \cite{feret:thesis}, we prove an authentication property \cite{BlanchetSAS02} in a cryptographic protocol \cite{Woo}   in the \spicalcul\ \cite{spi}.

\paragraph{Outline.}
We discuss related works in Sect.~\ref{related}.
We detail the contribution of this paper in Sect.~\ref{contribution}.
We give some examples in Sect.~\ref{examples}.
We give in Sect.~\ref{non-stan} a non-standard semantics for the \picalcul.
We define  both thread and step partitioning in Sect.~\ref{partition class}.
We derive a generic abstraction in Sect.~\ref{ai}.
We give an environment analysis in Sect.~\ref{environment}  
and a contents analysis in Sect.~\ref{occurrence}.

\section{Related works}
\label{related}
In this section, we discuss some related works. 
\subsection{Control flow analyses} 
Our analysis requires an accurate description of the potential interactions between the agents. Many type systems \cite{hennessy.riely:resource-access} and control flow analyses \cite{nielson:journal,nielson:concur98} propose a uniform description of these interactions in which recursive instances cannot be distinguished. In \cite{feret:jlap,feret:sas2000,feret:esop2002}, we proposed non-uniform control flow analyses, which distinguish between recursive instances of names. 
All these analyses abstract away the properties about concurrency.
\subsection{Groups}
Groups \cite{cardelli00secrecy,cardelli00ambient} are used in type system to prevent certain communications.
Recursive instances of groups are distinguished. The communication of a name outside the initial scope of its group is forbidden. 
On the contrary, our analysis computes relationship between the partition classes of interacting threads. So we can analyze systems where a name first exits the scope of the thread that had declared it and then returns inside this scope.  
\subsection{Numerical domains and concurrency}
Numeric analyses are widely used to analyze concurrency properties such as mutual exclusion and non-exhaustion of resources.
Disjunctive completion-based domains are used  in \cite{HJNN99ai} to count globally the components in \ambients\ and  in \cite{NiNi00popl,gori:aplas} to count the components inside each \ambient. 
These domains ignore the algebraic structure of numerical properties. 
Consequently, these analyses are exponential in time.
In \cite{feret:jlap,feret:getco2000}, we use affine equalities to count the threads of \picalcul\ systems  in polynomial time.
This analysis counts threads globally, regardless of their linkage. 
In the present paper, we  use information about the dynamic linkage of threads to gather threads in partition classes. Then we count the number of threads inside each partition class. Our approach is model-independent \cite{feret:thesis}.
Besides, we can detect and prove history-dependent and spatial-dependent properties (e.g.~see Ex.~\ref{content}).
\subsection{Behavioral types}
Behavioral types can express complex concurrency properties such as the absence of race conditions. 
But, in \cite{igarashi.kobayashi:generic-type}, some properties involving several names  cannot be checked because of the abstraction (e.g.~see Ex.~\ref{twonames}). 
The type system in \cite{rajamani.rehof:behavioral-module,rehofpopl02} can express and check more properties, but the type checking algorithm does not always terminate, whereas our inference algorithm  does in polynomial in time.
Moreover, 
 our occurrence counting and control flow analyses refine each other thanks to local trace partitioning. 
In Ex.~\ref{cfahelp}, we cannot analyze precisely mutual exclusion without the help of a precise control flow analysis.

\section{Contribution}
\label{contribution}
In this section, we describe the main contributions of this paper. 

This paper is a summary of the framework proposed in \cite[Chap.~10]{feret:thesis}. This framework is generic with respect to the model. In this paper,  we focus on systems that are written in the $\pi$-calculus. The main contributions of this paper are the following:
\begin{enumerate}
\item \emph{thread partitioning}: in this paper, we partition the threads of a mobile system according to some semantics criteria;
\item \emph{local trace partitioning}: then, we provide an extended labeled transition system in which each  computation step is annotated with information about the partition classes of the threads that interact; this  allows several analyses to share information about the partition classes of the threads that interact;
\item \emph{control flow analysis}: we refine existing analyses \cite{feret:sas2000,feret:esop2002,feret:jlap,feret:thesis} so as to take into account the constraints about the partition classes of the threads that interact;
\item \emph{content analysis}: we propose a new analysis to count the number of threads inside each partition class; this analysis is parametric with respect to a numerical domain (we use the same domain as in the occurrence counting analysis \cite{feret:getco2000,feret:jlap,feret:thesis} that counts the number of threads in the whole system).
\end{enumerate}

\section{Examples}
\label{examples}

In this section, we give some examples to motivate our framework. 

\subsection{Our running example}

First, we introduce an example that is easy to analyze: we  prove that there are never two simultaneous outputs over the same channel in a shared memory written in the \picalcul.  We give a manual proof in order to stress the  properties that are useful during the analysis. The goal of this example is just to understand how the analysis behaves: we use this example all along the paper.

We use  a version of the \picalcul\  inspired from
\cite{milner:polyadic,turner:thesis,cham}.
Let $\Names$ be  an infinite set of variables and $\Labels$ be a finite set of labels. 
Let $c,x,y\in \Names$ be some variables, $l\in\Labels$ be a label, 
and $\overline{x}\in \Names^{\ast}$ be a tuple of variables.
The agent $(P\;\concu\; Q)$ denotes the parallel composition of two agents $P$ and $Q$. It performs $P$ and $Q$ simultaneously. 
The agent $(\nuu x)P$ binds the variable $x$ to a fresh channel name in $P$.
The agent $\emptypro$ does nothing (it is usually omitted).
The agent $c\eme^l[\overline{x}].P$  sends a message (i.e.~a tuple of channel names) via the channel the name to which the variable $c$ is bound. 
The agent $c\rec^l[\overline{x}].P$ waits for a message on the channel  to which the variable $c$ is bound, and binds the tuple $\overline{x}$ of distinct variables to the received  names. The agent $\repli c\rec^l[\overline{x}].P$ is a \emph{resource} which replicates itself when  receiving messages.
Name restriction $(\nuu x)P$ and message reception $c\rec^l[\overline{x}].P$ or $\repli c\rec^l[\overline{x}].P$ are the only variable binders.
We denote by $\FN(P)$ the set of the variables that are free in $P$. 
Labels help in locating syntactic components. 
Moreover, the notation $\bang{l} P$ stands for $(\nuu\cbang_l)(\cbang_l\eme^{l}[] \;\concu \repli\cbang_l\rec^{l'}[].(\cbang_l\eme^{l''}[]  \;|\;P))$ where $l$, $l'$, and $l''$ are fresh labels, and $\cbang_l \not\in\FN(P)$: it denotes an unbounded number of concurrent instances of $P$.

{
\begin{example}[a shared memory]
\label{memory}
A shared memory with dynamic allocation of cells may be described  
in the \picalcul\ as follows:
\begin{tabbing}
$(\nuu\ccreate)(\nuu\cnull)$\\
$($\;$\repli$\=$\ccreate\rec^{\memlaba}[\textit{address}].(\nuu \ccell)(\nuu \cread)(\nuu \cwrite)$\\
\>$($\=\;$\ccell\eme^{\memlabb}[\cnull]\;\concu\; \textit{address}\eme^{\memlabc}[\cread,\cwrite]\;$\\
\>$\concu\; \repli\!\cread\rec^{\memlabd}[\port].\ccell\rec^{\memlabe}[\dataread].(\ccell\eme^{\memlabf}[\dataread]\;\concu\;\port\eme^{\memlabg}[\dataread])$\\
\>$\concu\;\repli\!\cwrite\rec^{\memlabh}[\datawrite,\cack].\ccell\rec^{\memlabi}[v].(\ccell\eme^{\memlabj}[\datawrite]\;\concu\;\cack\eme^{\memlabk}[]))$\\
$\concu\;\bang{\memlabl} 
(\nuu\addinit)$$\ccreate\eme^{\memlabm}[\addinit].\addinit\rec^{\memlabn}[\textit{read},\textit{write}].$\\
\>$(\;\bang{\memlabo} (\nuu \addread)\textit{read}\eme^{\memlabp}[\addread].\addread\rec^{\memlabq}[x]$\\
\>$\concu\;\bang{\memlabr} (\nuu \cdata)(\nuu \addwrite)\textit{write}\eme^{\memlabs}[\cdata,\addwrite].\addwrite\rec^{\memlabt}[]))$
\end{tabbing}
Whenever a message is sent via the channel name declared by the restriction $(\nuu \ccreate)$ \pps{\memlaba}, a memory cell is allocated.
Three names are introduced. The name $\ccell$ encodes the contents of the memory cell: the contents of the cell are always output once over the channel named $\ccell$ (the name $\cnull$ denotes the initialization value); the names $\cread$ and $\cwrite$ encode respectively the capability to read and to overwrite the contents of the cell. 
The client is given the capability to interact with the cell \pps{\memlabc}.
The memory can deal with an unbounded number of read \pps{\memlabd} and write \pps{\memlabh} requests. 
A read request requires a return address to which the contents of the cell are forwarded (please note that we copy the contents of the cell once, so as not to lose them). 
A write request requires two arguments, the new contents and an acknowledgment address: the cell contents  are first removed and then replaced with  the new contents, the acknowledgment controls client requests sequentiality. 
An unbounded number of clients are created \pps{\memlabk}.
Each client creates a cell and performs an arbitrary number of read \pps{\memlabo} and write \pps{\memlabr} requests.

We want to prove that there is never more than one simultaneous output on any channel $c$ opened by an instance of the restriction $(\nuu \ccell)$.
First, we propose a manual proof to give intuitions about our framework.
The analysis 
in this paper discovers this property automatically.
Let us denote by $\mathcal{M}$ the set of the names introduced by an instance of the restriction $(\nuu \ccell)$, we will prove that at any configuration of the system and any name $c\in \mathcal{M}$: there is either no thread, or exactly one output (at program point \pp{\memlabb}, \pp{\memlabf}, or \pp{\memlabj}) on the corresponding channel.
For any configuration $C$ and any name $c\in\mathcal{M}$, we define $y(C,c)$ as $0$ whenever the name $c$ has not been allocated yet, and as $1$ otherwise. We denote by $x_i(C,c)$ the number of threads at program point $i$ that operate on the channel named $c$. 
Now, we prove by induction over the history of the system that $x_2(C,c)+x_6(C,c)+x_{\memlabj}(C,c)-y(C,c)=0$. At the beginning of the system, we have, for any $c\in\mathcal{M}$, $x_2(C,c)=x_6(C,c)=x_{\memlabj}(C,c)=y(C,c)=0$, so the property holds. 
When two threads at program points \pp{\memlaba} and \pp{\memlabm} interact, a fresh name $c$ is allocated. Since this name is fresh, we have, before the interaction, $x_2(C,c)=x_6(C,c)=x_{\memlabj}(C,c)=y(C,c)=0$; after the interaction,  we have $x_2(C,c)=y(C,c)=1$ and $x_6(C,c)=x_{\memlabj}(C,c)=0$. This way, the property still holds. 
We now consider an interaction  between a thread $t_i$ at program point $i\in \{5;9\}$ and a thread $t_j$ at program point $j\in\{2;6;10\}$; 
this interaction launches a thread at program point $i+1$.
We consider  several cases according to the relationships among the channels on which these three threads operate. 
There are $5$ cases: they may operate on the same channel, on two distinct channels (three cases), or on three distinct channels. 
We use a control flow analysis to detect which cases are possible: we detect that the only possible case is the case where the three threads operate on the same channel $c$. 
During the transition, $x_{j}$ is decremented and $x_{i+1}$ is incremented ($x_j$ is not changed when $j=i+1$), so the property of interest  still holds.\boxexample
\end{example}}

\subsection{More complex examples}

In this section, 
we describe more complex examples in order to illustrate some difficulties that can be tacked by our analysis.

{
\begin{example}[related names]
Our analysis can  abstract the usage of several names together. We consider the following system (adapted from \cite{igarashi.kobayashi:generic-type}) in the \picalcul:
\label{twonames}
\begin{tabbing}
$(\nuu\alloc)$
$($$\repli$\=$\alloc\rec^{\twolaba}[\addone,\addtwo,\addthree].
(\nuu \clock)(\nuu \cm)(\nuu \crr)$\\
\hspace*{1cm}$(\clock\eme^{\twolabb}[]\;\concu\;\addone\eme^{\twolabc}[\clock]\;\concu\; \addtwo\eme^{\twolabd}[\cm]\;\concu\; \addthree\eme^{\twolabe}[\crr]\;\concu\;\bang{\twolabf}\cm\rec^{\twolabg}[].\crr\eme^{\twolabh}[])$\\
$\concu \bang{\twolabi}\!(\nuu \argone)(\nuu \argtwo)(\nuu \argthree)(\alloc\eme^{\twolabj}[\argone,\argtwo,\argthree].$\\
\hspace*{1cm}$\argone\rec^{\twolabk}[l].\argtwo\rec^{\twolabl}[m].\argthree\rec^{\twolabm}[r].\bang{\twolabn}l\rec^{\twolabo}[].m\eme^{\twolabp}[].r\rec^{\twolabq}[].l\eme^{\twolabr}[]))$
\end{tabbing}
The server \pps{\twolaba} creates several objects. Each object is made of a lock \clock, a method \cm, and a return address \crr. 
Each session \pps{\twolabo} consists in locking the method, calling the method, receiving the returned value (which is abstracted away),  
and then releasing the lock. 
There is an unbounded number of clients \pps{\twolabi}. 
Each one creates an object \pps{\twolabj}, 
receives the lock, the method, and the returned address during three channeled communications, and performs an arbitrary number of sessions \pps{\twolabn}. 
We partition\footnote{This partitioning is made possible in the non-standard semantics where each name is tagged with the identifier of the thread that has declared it (see Sect.~{\ref{non-stan}}).}
 the threads according to the recursive instance of the resource that has declared the name of the channel on which each thread operates.
Our analysis detects and proves that there can never be more than one simultaneous call of the same method.  This result is beyond the reach of \cite{igarashi.kobayashi:generic-type}, because the names \clock, \cm, and \crr\  are not communicated during a single communication.
\boxexample
\end{example}}

\begin{example}[control flow dependence]
\label{cfahelp}
We now illustrate the importance of the control flow analysis.
We describe a doubly-linked list of cells as follows:

\begin{tabbing}
$(\nuu\crec)$\=$(\nuu\llinkint{_0})(\nuu\clinkint{_0})$\=$(\nuu\rlinkint{_0})(\nuu\cset)$\\
$(\crec\eme^{\cfalaba}\celllinkint{_0}$\\
$\concu
\repli\crec\rec^{\cfalabc}\celllinkvar{_n}.
    (\nuu\llinkint{_{n+1}})
    (\nuu\clinkint{_{n+1}})(\nuu\rlinkint{_{n+1}})$\\
\>$(\crec\eme^{\cfalabd}\celllinkint{_{n+1}}\;\concu\;\clinkvar{_n}\eme^{\cfalabe}[]\;\concu\; \bang{\cfalabf}\cset\eme^{\cfalabg}\celllinkvar{_n}$\\
\>$\concu\;\bang{\cfalabh}\llinkint{_{n+1}}\eme^{\cfalabi}\celllinkvar{_n}
       \;\concu\; \bang{\cfalabj}\rlinkvar{_n}\eme^{\cfalabk}\celllinkint{_{n+1}})$\\ 
$\concu\repli\cset\rec^{\cfalabl}\celllinkvar{}.\rlinkvar{}\rec^{\cfalabm}\celllinkvar{'}.\llinkvar{'}\rec^{\cfalabn}\celllinkvar{''}.\clinkvar{}\rec^{\cfalabo}[].\clinkvar{''}\eme^{\cfalabp}[])$
\end{tabbing}
Each cell is encoded by three names: the name $\llinkint{_n}$ encodes a backward pointer to the previous cell, the name $\clinkint{_n}$ encodes the cell address (the contents are abstracted away), and  the name $\rlinkint{_n}$ encodes a forward pointer to the next cell.
Each cell is output on the channel named $\cset$ \pps{\cfalabg}. 
Then, at program point ${\bf {\cfalabl}}$, we pick a cell.
We collect its address $\clinkvar{}$, we follow the forward pointer, 
then we follow the backward pointer, 
and we collect the address $\clinkvar{''}$ of the reached cell. 
The control flow analysis \cite{feret:sas2000} detects that the addresses
 $\clinkvar{}$ and $\clinkvar{''}$ are the same. This information is passed to the occurrence counting domain thanks to the local trace partitioning. Thus, we prove automatically that there is no simultaneous outputs over an instance of a channel named $\clinkint{_n}$.

It may look a bit curious to use two variables for the same name. 
But, these kinds of things are common in automatically generated systems. 
With a more general point of view, this difficulty is similar to the problem of aliasing in data structures. \boxexample
\end{example}

{
\begin{example}[a $2$-semaphore]
Our analysis is not limited to the detection of mutual exclusion.
In the following example: 
\begin{equation*}
\bang{\semlaba}(\nuu \cana)(\cana\eme^{\semlabb}[] \;|\; \cana\eme^{\semlabc}[] \;|\; \repli \cana\rec^{\semlabd}[].\cana\eme^{\semlabe}[]),
\end{equation*}
 our analysis detects automatically that 
there are never more than two simultaneous outputs over an instance of the channel $\cana$.
Besides, our analysis detects and proves the number of simultaneous outputs without requiring a bound on the number of copies that have to be distinguished by the analyzer. 
\boxexample
\end{example}}

{\begin{example}[synchronous communications]
Content analysis can also refine the control flow analysis.
In the following system:
\begin{equation*}
(\bang{\synlaba}(\nuu \cana)(\nuu \canb)(\nuu \canc)
(\cana\eme^{\synlabb}[\canb].\cana\rec^{\synlabc}[\varu].\varu\eme^\synlabd[\varu]\;|\;
\cana\rec^{\synlabe}[\varv].\cana\eme^{\synlabf}[\canc].\varv\eme^\synlabg[\varv])),
\end{equation*}
the content analysis detects that, for each instance, the thread at the program point \pp{\synlabb} (resp.~\pp{\synlabf}) and the thread at the program point \pp{\synlabc} (resp.~\pp{\synlabe}) are in mutual exclusion. 
The control flow analysis uses this information to prove that the variable $\varu$ (resp.~$\varv$) can only be bound to a channel opened by the restriction $(\nuu  \canc)$ (resp.~$(\nuu \canb)$).
\boxexample\end{example}}

\subsection{An example in mobile \ambients}

Our last example is written in another process calculus to illustrate that our framework is generic.

In mobile \ambients\ \cite{ambients}, 
a system is described by a hierarchy of named sites $n^l[P]$, called \ambients\ ($n$ is a name, $l$ is a label, and $P$ is a process).
\Ambients\ may contain some other \ambients\  and some agents $\textit{in}^l n.P$/$\textit{out}^l.P$/$\textit{open}^l n.P$ that provide them the capability to move in the hierarchy of \ambients\ or to open  some \ambients\ 
 (when an \ambient\ opens another one, the former \ambient\ gets the contents of the later). These interactions are controlled both by \ambient\ names (the name of the  target \ambient\ and the name occurring in the capability must be the same) and by spatial constraints. 
These interactions are described by the following reduction rules:
 $m^i[\textit{in}^k n.P\;|Q]\;|\;n^j[R] \stackrel{\scriptscriptstyle i,j,k}{\scriptstyle \longrightarrow} n^j[ m^i[P\;|\;Q]\;|\;R]$,  $n^j[ m^i[\textit{out}^k n.P\;|\;Q]\;|\;R]\stackrel{{\scriptscriptstyle i,j,k}}{\scriptstyle \longrightarrow} m^i[P\;|Q]\;|\;n^j[R]$, and 
$\textit{open}^i n.P \;\concu\; n^j[Q] \stackrel{{\scriptscriptstyle i,j}}{\scriptstyle \longrightarrow}  P\;|\;Q$.
\Ambients\ are also fitted with communication primitives: the agent  $(x)^l.P$ waits for a message (that can be either a name or a capability path), whereas the agent $\langle y\rangle^l$ sends a message. 
These communications are not channeled because 
threads can communicate only when they are in a same \ambient.
Both \ambients\ and \ambient\ names can be created dynamically. As in the $\pi$-calculus, we use guarded replication: the agent $!(x)^l.P$ duplicates itself when receiving a message and the agent $!\textit{open}^l n.P$ duplicates itself when opening another \ambient.
\begin{example}[the contents of an \ambient]
\label{content}
A \emph{client-server} protocol  may be described in the \ambient-calculus as follows:
\begin{tabbing}
$(\nuu \make)(\nuu \server)(\nuu \duplicate)(\nuu \instance)$$(\nuu \client)$\\
$($\=$\mathrm{server}^\amblaba$\=$[$$\;!\textit{open}^{\amblabb} \duplicate.0\;$\\
\>\>$\concu\;!(\ambk)$\=${}^{\amblabc}.\instance^{\amblabd}[\textit{in}{}^{\amblabe} \ambk.\textit{out}^{\amblabf} \server.\textit{in}^{\amblabg}\client.0]]$\\
$\concu\;\client^{\amblabh}$\=$[\;!(\ambx)^{\amblabi}.$\=$((\nuu \ambp)$$\ambp^{\amblabj}$\=$[\;$\=$\textit{out}^{\amblabk} \client.0\;\concu\;\textit{open}^{\amblabl} \instance.0\;$\\
\>\>\>$\concu\;\textit{in}^{\amblabm} \server.\duplicate^{\amblabn}[\textit{out}^{\amblabo} \ambp.<\ambp>^{\amblabp}]]$\\
\>\>$\concu\;<\make>^{\amblabq}$
$)$
$\;\concu\;<\make>^{\amblabr}$
$])$
\end{tabbing}
In this protocol, some packets are created \pps{\amblabj}. They are initially located in the \ambient\ ${\bf{\bfclient^{\amblabh}[\bullet]}}$. Each packet is identified by a fresh name $\bfp$. The packet contains some routing information to enter the \ambient\ ${\bf{\bfserver^{\amblaba}[\bullet]}}$. Once inside the server \ambient, the packet expels an \ambient\ ${\bf{\bfduplicate^{\amblabn}[\bullet]}}$ in order to communicate the name of the packet to the server.
The server may open this \ambient\ \pps{\amblabb}, receive the name of the packet \pps{\amblabc}, and create an \ambient\ \pps{\amblabd} that enters the packet. Then the packet opens \pps{\amblabl} this \ambient\ to receive the capability to return inside the \ambient\ ${\bf{\bfclient^{\amblabh}[\bullet]}}$.
In this example, we abstract away what is happening to the packet while it is in the server domain.

We partition the threads (both agents and \ambients) according to their location and the location of their surrounding \ambient. 
Then, we count the number of threads inside each class of the partition.
Our analysis discovers   the contents of the packet according to its position in the network. 
For instance, we detect that whenever the packet is inside the \ambient\ ${\bf \bfclient^{\amblabh}[\bullet]}$: 
it contains only threads at the program points \pp{\amblabd}, \pp{\amblabk}, \pp{\amblabl}, and \pp{\amblabm}; 
moreover, either there is exactly one thread 
at each program point \pp{\amblabk}, \pp{\amblabl}, and \pp{\amblabm}, or 
no threads at these three program points.
Similar information are inferred for the other potential locations of the packet.
We notice that our analysis loses all information about the number of threads at program point \pp{\amblabd}, because it cannot infer that for a given packet, only one instance can receive the name of the packet. But since we detect that only one can be opened, this has no influence on the inference of the other properties.
As in Ex.~\ref{memory}, we require an abstraction of the local history of each packet to reach this accuracy level:
we count the number $y_{\lambda}$ of each kind $\lambda$ of transition,
 we also consider the variables $z_{\lambda}$ that are defined as $z_{\lambda}=0$ if $y_{\lambda}=0$, and $z_{\lambda}=1$ otherwise. \boxexample
\end{example}

\section{Non-standard semantics}
\label{non-stan}

To prove the properties that interest us, we need to distinguish recursive instances of threads. 
Standard semantics are not convenient, because the $\alpha$-conversion breaks the relations between the threads and the name of the channels that they open. In this section, we recall a non-standard semantics \cite{feret:sas2000,feret:jlap,feret:thesis}. This semantics is more concrete: each thread is annotated with information about both its history and the history of the names that it handles.

\subsection{Notations}
We consider a closed mobile system $\System$ (i.e.~$\FN(\System)=\emptyset$) in the \picalcul.
We may assume  that each variable is bound  exactly once in the system (either by a name restriction or by an input).
We may also assume that syntactic components  are labeled with distinct labels.
For any label $l\in\prolabel$, we denote by $\component(l)$  the subprocess the first action of which is labeled with  $l$.
We define $\type(l)$ as $\inputm$ if  $\component(l)$ matches $c\rec^l[x_1,\ldots,x_n].Q$, as $\outputm$ if  $\component(l)$ matches $c\eme^l[x_1,\ldots,x_n].Q$, and as $\fetch$ if  $\component(l)$ matches $\repli c\rec^l[x_1,\ldots,x_n].Q$.
Besides, with the same notations, we define $\canal(l)\bydef c$, $\args(l)\bydef[x_1,\ldots,x_n]$, and $\cont(l)\bydef Q$. 
For any process $P$, we define the set $\Extract(P)$ of the labels of the threads that are launched in $P$, 
$\Extract(P \;\concu\; Q)  \bydef   \Extract(P) \cup \Extract(Q)$, 
$\Extract((\mathbf{\nuu}x)P) \bydef  \Extract(P)$, 
$\Extract(\mathbf{0})  \bydef  \emptyset$, and 
$\Extract(c\eme^l[x_1,\ldots,x_n].P) \bydef \Extract(c\rec^l[x_1,\ldots,x_n].P) \bydef
\Extract(\repli c\rec^l[x_1,\ldots,x_n].P) \bydef \{l\}$.
For any label $l$, we denote by $\interface(l)$ the set of the variables that are free  in the threads at program point $l$.
Thus, we define $\interface(l)$ as $\FN(\component(l))$.

\begin{figure*}[t]
\subfigure[Non-standard initial configuration.]
{\begin{minipage}[c]{\linewidth}
\begin{equation*}
\initstan\bydef \launch(S,\varepsilon,\emptyset)
\end{equation*}
\end{minipage}}

\subfigure[Non-standard transition system.]
{\begin{minipage}[c]{\linewidth}
\begin{equation*}
\fracbis{\displaystyle 
\begin{cases}
\Erec(\canal(\lrec))=\Eeme(\canal(\leme)),\cr
\type(\lrec)=\inputm,\;\type(\leme)=\outputm,\cr
[y_1,\ldots,y_n]\bydef \args(\lrec),\;[x_1,\ldots,x_n]\bydef \args(\leme),\hspace*{-0.2cm}\cr
\textit{Ct}_{\rec}\bydef\launch(\cont(\lrec),\idrec,\Erec[y_k\mapsto \Eeme(x_k)]),\;\cr
\textit{Ct}_{\eme}\bydef\launch(\cont(\leme),\ideme,\Eeme),\cr\end{cases}\hspace*{-4pt}}
{\displaystyle C\cup\{\ttuplerec;\ttupleeme\} 
\nonstan{(\lrec,\leme)} (C \cup \textit{Ct}_\rec \cup \textit{Ct}_\eme)}
\end{equation*}

\vspace*{5mm}

\begin{equation*}
 \fracbis{\displaystyle \begin{cases}
\Erec(\canal(\lrec))=\Eeme(\canal(\leme)),\cr
\type(\lrec)=\fetch,\;\type(\leme)=\outputm,\cr
[y_1,\ldots,y_n]\bydef \args(\lrec),\;[x_1,\ldots,x_n]\bydef \args(\leme),\hspace*{-0.2cm}\cr
\textit{Ct}_{\rec}\bydef\launch(\cont(\lrec),\leme.\ideme,
\Erec[y_k\mapsto E_!(x_k)]),\cr
\textit{Ct}_{\eme}\bydef\launch(\cont(\leme),\ideme,\Eeme),\cr
\end{cases}\hspace*{-4pt}}
{\displaystyle C\cup\{\ttuplerec;\ttupleeme\}
\nonstan{(\lrec,\leme)} 
(C \cup \{\ttuplerec\} 
\cup \textit{Ct}_{\rec} \cup \textit{Ct}_{\eme})}
\end{equation*}
\end{minipage}}
\caption{Non-standard semantics.}
\label{non-standard semantics}
\end{figure*}

\subsection{Semantics}
We define a non-standard semantics in which both threads and channel names are tagged with the history of their creation. 
History markers $\textit{id}\in \Markers$ are sequences of labels in $\Labels$.
Markers encode the history of the replications which have led to the creation of thread instances. 
The markers of initial threads are $\varepsilon$. 
When a computation step does not involve fetching a resource,
markers are just passed to the continuations; when a resource is fetched, the new instance is tagged with  $\leme.\ideme$ where $\leme$ and $\ideme$ are respectively  the label and the marker of the output thread. 

Then, we stamp each  name with  the marker of the thread  which has declared it. Thus, a channel name is a pair $(x,\textit{id})$ composed of a variable $x\in\Names$ and a marker $\textit{id}\in\Markers$, which means  that this is the name of the channel that has been opened by the restriction $(\nuu x)$ of a thread tagged with the marker  $\textit{id}$.

A \emph{configuration} of the system $\System$ is  a set of \emph{thread instances}. 
Each thread instance is a $3$-tuple composed of a \emph{label} $l\in\Labels$  that denotes a syntactic component, an unambiguous \emph{marker} $\textit{id}\in\Markers$, and an \emph{environment} $E\in \interface(l)\rightarrow \Names\times \Markers$ which specifies the channel names to which free variables are bound.
Thread instances are created at the beginning of the computation and when agents interact. 
The function  $\launch$ applied to a subprocess, a marker, and an environment, collects all the threads that are spawned when a continuation is launched: we set 
 $\launch(P,\textit{id},E)\bydef\{(l,\textit{id},E_l)\;|\;l\in \Extract(P)\}$, where, $E_l\in\interface(l)\rightarrow \Names\times \Markers$ maps 
any $x\in \interface(l)\cap \textit{Dom}(E)$ to $E(x)$,  
and any $x\in\interface(l)\setminus\textit{Dom}(E)$ to $(x,\textit{id})$.
This simulates name restriction by binding any new  variable $x$ to the name of the channel opened by the restriction $(\nuu x)$ of a thread the marker of which is $\textit{id}$.
The initial state and computation rules are given in Fig. \ref{non-standard semantics}. The correspondence between the non-standard and the usual semantics  is proved   in \cite{feret:jlap,feret:thesis}.

\begin{example}[\cnonunu]
We apply our non-standard semantics with our shared memory example (see.~Ex.~\ref{memory}). We obtain the initial state $C_0 = \{t_1;t_2;t_3\}$ where:
\begin{equation*}
\begin{array}{l}
t_1 = (\memlaba,\varepsilon,[\ccreate\mapsto(\ccreate,\varepsilon),\cnull\mapsto(\cnull,\varepsilon)]),\cr
t_2 = (\memlabl,\varepsilon,[\crec_{\memlabl}\mapsto(\crec_{\memlabl},\varepsilon)]), \text{ and} \cr
t_3= (\memlabl',\varepsilon,[\ccreate\mapsto(\ccreate,\varepsilon),\crec_{\memlabl}\mapsto(\crec_{\memlabl},\varepsilon)]).\cr
\end{array}
\end{equation*}
The thread $t_1$ is a resource that can allocate memory cells, the thread $t_2$ can interact with the thread $t_3$ to create clients recursively. Since these three  threads  are in the initial state, their thread marker is $\varepsilon$. 

We create a first client by making the threads $t_2$ and $t_3$ interact. 
We obtain the state $C_1 = \{t_1;t_3;t_4;t_5\}$, where:
\begin{equation*}
\begin{array}{l}
 t_4=(\memlabl'',\memlabl,[\crec_{\memlabl}\mapsto(\crec_{\memlabl},\varepsilon)]) \text{ and }\cr
t_5=(\memlabm,\memlabl,[\ccreate\mapsto(\ccreate,\varepsilon),\addinit\mapsto(\addinit,\memlabl)]).\cr
\end{array}
\end{equation*}
The thread $t_4$ 
allows the creation of further clients 
and the thread $t_5$ describes a client that can allocate a memory cell. 
We create a second client by making the threads $t_4$ and $t_3$ interact. 
We get the state $C_2 = \{t_1;t_3;t_5;t_6;t_7\}$, where:
\begin{equation*}
\begin{array}{l}
t_6=(\memlabl'',\memlabl''.\memlabl,[\crec_{\memlabl}\mapsto(\crec_{\memlabl},\varepsilon)]) \text{ and } \cr
t_7=(\memlabm,\memlabl''.\memlabl,[\ccreate\mapsto(\ccreate,\varepsilon),\addinit\mapsto(\addinit,\memlabl''.\memlabl)]).
\end{array}
\end{equation*}
The thread $t_6$ 
allows the creation of further clients 
and the thread $t_7$  describes the second  client. 
Both clients are identified by their thread markers $\memlabl$ and $\memlabl''.\memlabl$. Besides, the link between threads and the channel names that they handle is explicit: the thread $t_5$ can operate on the name $(\addinit,\memlabl)$, whereas the thread $t_7$ can operate on the name $(\addinit,\memlabl''.\memlabl)$.
\boxexample\end{example}

\section{Thread partitioning and trace partitioning}
\label{partition class}

In this section, we first partition the threads of the system in several partition classes. Then we partition computation steps according to some relations about the threads that are involved. As a result, we obtain an extended labeled transition system.

\subsection{Thread partitioning}
Let $B$ be a finite set of keys. 
Our analysis is parameterized by  a function $\giveindex$ mapping each program point label $l$ to a function in $B\rightarrow \interface(l)$. 
Then, we partition the threads $t=(l,\textit{id},E)$ in a configuration according to the value $E(\giveindex(l)(b))$ of the variable $\giveindex(l)(b)$ for each key $b\in B$. 
We can also partition threads according to the markers of their names (we focus on partitioning according to full names to simplify the presentation.).
For example, to prove the absence of race conditions, 
we gather the threads that operate on the same channel (we define  $B$ as $\{b\}$ and $\giveindex(l)(b)$ as $\canal(l)$).
In \ambients, we partition threads in accordance with their location and the location of their surrounding \ambient\ (thus, $B$ contains two keys). 
We know that: whenever two threads are in the  same \ambient, the location of their surrounding \ambient\ is the same (partitioning the threads also 
according to the location of their surrounding \ambient\ allows for a more precise  partitioning at the abstract level). 
We denote by $B_s\subseteq B$ a  set of keys, such that:
for any configuration $C$, for any threads $t_1=(p_1,\textit{id}_1,E_1)$ and $t_2=(p_2,\textit{id}_2,E_2)$ in the configuration $C$, if,
for any $b\in B_s$, $E_1(\giveindex(p_1)(b))=E_2(\giveindex(p_2)(b))$, then,  for any $b\in B$, $E_1(\giveindex(p_1)(b))=E_2(\giveindex(p_2)(b))$.
This implication will be useful whenever we know that two threads are in the same configuration but in distinct partition classes.

Each partition class is identified by a function $f\in B\rightarrow \Names\times \Markers$, called \emph{computation unit}.
We denote by $\computationunit$ the set $B\rightarrow \Names\times \Markers$ of all computation units.
There may be an unbounded number of computation units. 
We gather them into a finite set of abstract computation units by abstracting away the information about markers: 
we define the set $\abstractunit$ of abstract computation units as $B\rightarrow \Names$. The abstraction function $\piunit$ maps each computation unit  $[(b\in B)\mapsto (l_b,\textit{id}_b)]\in \computationunit$ to the abstract one $[b\mapsto l_b]\in \abstractunit$.

\subsection{Local trace partitioning}
\newcommand{\vartrans}{\mathcal{T}(\lrec,\leme)}
We consider a computation step $\tau=(C\nonstan{(\lrec,\leme)} C')$. We denote by $\trec=\ttuplerec$ and by $\teme=\ttupleeme$ the threads that interact in the computation step $\tau$. 
The thread $\trec$ launches one thread for each label  $l$ in the set $\beta(\cont(\lrec))$ and the thread $\teme$ launches  one thread for each label $l$ in the set $\beta(\cont(\leme))$. We denote by $n_{\rec}$ the cardinal of the set $\beta(\cont(\lrec))$ and by $n_{\eme}$ the cardinal of the set $\beta(\cont(\leme))$. 
Thus, the computation step $\tau$ involves $2+n_{\rec}+n_{\eme}$ threads. 
Each of these threads is denoted by a pair $(l,\diamond)\in\prolabel\times \{\rec;\eme\}$ where $l$ is the label of the thread program point and $\diamond$ is equal to $\rec$ when this thread is related to the input  thread or to $\eme$ when this thread is related to the output thread. This way, we denote by $\vartrans$
 the set $\{(l,\diamond)\;|\;\diamond\in\{\rec;\eme\},\;l\in\{l_\diamond\}\cup\beta(\cont(l_\diamond))\}$. 
\newcommand{\abstractcontext}{\textsc{context}}
\newcommand{\classes}{(\vartrans)_{\sim}}
\newcommand{\classede}[1]{[#1]_{\sim}}

To get a more precise analysis, 
we partition the set of computation steps according to some properties about the computation units of the threads that are involved in these computation steps.
We denote by $\abstractcontext(\lrec,\leme)$ the set of pairs $(\sim,A)$ such that $\sim$ is an equivalence relation\footnote{Given an equivalence relation $\sim$ over a set $A$, $[a]_{\sim}$ denotes the equivalence class $\{b\in A\;|\;a\sim b\}$ of  $a$ and $A_{\sim}$ denotes the set $\{[a]_\sim\;|\;a\in A\}\}$ of equivalence classes.}  in $\wp(\vartrans^2)$ that relates the threads that share the same computation unit  and $A\in \classes \rightarrow \abstractunit$ maps each equivalence class to its abstract computation unit.
Intuitively, the relation $(l_1,\diamond_1)\sim (l_2,\diamond_2)$ means that the thread denoted by the pair $(l_1,\diamond_1)$ and the thread denoted by the pair $(l_2,\diamond_2)$ are in the same computation unit; moreover, $A([(l,\diamond)]_{\sim})$ is the abstract computation unit of the thread denoted by the pair $(l,\diamond)$. 
More formally, we denote by $\textit{unit}_{\tau}$ the function which maps any pair $(l,\diamond)\in\vartrans$ to the computation unit of the thread denoted by the pair $(l,\diamond)$.
Then, we define the abstraction function $\alpha_{\textsc{step}}$ which  maps each computation step to its partition case as: 
$\alpha_{\textsc{step}}(\tau) \bydef (\sim,[[a]_{\sim} \mapsto \piunit(\textit{unit}_{\tau}(a))])$, 
where $\sim$ is defined as $a\sim  b$ if and only if $\textit{unit}_{\tau}(a)=\textit{unit}_{\tau}(b)$.

\section{Abstraction}
\label{ai}
In this section, we use the abstract interpretation framework \cite{c:thesis,cc:popl77} to design a generic abstraction of transition systems. 

\subsection{Reachable states}
We denote  by $\states$ the set  of all configurations, by $\extendedSigma$ the set of pairs 
in $\bigcup_{\lambda\in\translabel} \{\lambda\}\times \abstractcontext(\lambda)$, and, for any finite set $V\subseteq \Names$ of variables, by $\env{V}\bydef\Env{V}$ the set of  marker/environment (over $V$) pairs.
We are interested in $\Coll$, the set  of all configurations that are reachable through a finite computation sequence.
The set $\Coll$ is the least fixpoint of the $\cup$-complete endomorphism $\mathbb{F}$ on the complete lattice $\wp(\states)$, where $\mathbb{F}$ is defined as $[X \mapsto \initstan \cup \{\overline{C}\in\states\;|\; \exists C\in
X,\;\exists \lambda\in\extendedSigma,\;C\nonstan{\lambda} \overline{C} \}]$.
This least fixpoint is usually not decidable, so 
 we use a relaxed version of the abstract interpretation framework
\cite{cc:frameworks}  to compute a sound---but not necessarily complete---approximation of it.

\subsection{Generic abstraction}
We choose an \emph{abstract domain}, which is  a set of abstract symbolic properties about configurations.
It captures  the properties of interest and abstracts away the other properties. Each abstract property is mapped to the set of the concrete elements which satisfy this property by a  concretization map $\conc{}\in\adom{}\rightarrow \wp(\states)$. The abstract domain is fitted with several primitives to handle its elements. 
An abstract union $\aunion{}\in \wp_{\textrm{finite}}(\adom{})\rightarrow \adom{}$ gathers the information described by several abstract elements. It satisfies: $\forall a^\sharp \in A^\sharp,\;\conc{}(a^\sharp) \subseteq \conc{}(\aunion{}(A^\sharp))$. 
We also need an abstraction $\ainit{} \in \adom{}$ of the initial configuration (i.e.~$\initstan \in \conc{}(\ainit{})$).
To simulate computation steps in the abstract, we introduce an abstract operator $\abst{}\in \adom{}\times\ExtendedSigma \rightarrow\adom{}$.  
This operator partitions each transition into several sub-cases: given an abstract property $C^\sharp\in\adom{}$ and a sub-case $\overline{\lambda}=(\lambda,\textit{context})\in\ExtendedSigma$, the set $\conc{}(\abst{}(C^{\sharp},\overline{\lambda}))$ contains all the states $\overline{C}\in\states$ that are reachable from any state $C \in \conc{}(C^\sharp)$ by a computation step $\tau = { C\nonstan{\lambda} \overline{C}}$ such that $\textit{context}=\alpha_{\textsc{step}}(\tau)$.
An abstract element $\abot{}$ such that $\conc{}(\abot{})=\emptyset$  provides the basis for our abstract iteration. Finally, we use a widening operator $\wid{}\;:\;\adom{}\times\adom{} \rightarrow \adom{}$ to ensure the termination of our analysis. It satisfies $\forall C_1^\sharp,\;C_2^\sharp \in \adom{},\;\conc{}(C_1^\sharp) \subseteq \conc{}(C_1^\sharp \wid{} C_2^\sharp)\text{ and } \conc{}(C_2^\sharp) \subseteq \conc{}(C_1^\sharp \wid{} C_2^\sharp)$; moreover,  
for any sequence $(C_n)\in{\adom{}}{}^\mathbb{N}$, the sequence $(C_n^{\wid{}})$ that is defined by $C_0^{\wid{}}\bydef C_0$ and $C_{n+1}^{\wid{}}\bydef C_n^{\wid{}}\wid{}C_{n+1}$ for any $n\geq 0$, is ultimately  stationary.
We do not use narrowing because, we iterate only functions $f\in\adom{}\rightarrow \adom{}$ that satisfy: 
$\conc{}(a)\subseteq \conc{}(f(a))$.
\begin{definition}
Any tuple $\tuple{}$ that satisfies these assumptions is called an abstraction.\end{definition}

Given an abstraction $\abstraction{}=\tuple{}$, we define the abstract counterpart $\counterpart{}$ of the function 
$\mathbb{F}$ as the function that maps any abstract element $C^{\sharp}\in\adom{}$ to the abstract element $\aunion{}
(\{\abst{}(C^{\sharp},\overline{\lambda}) \;|\; \overline{\lambda}\in\ExtendedSigma\}\cup\{\ainit{}\})$.
The function $\counterpart{}$ satisfies the soundness condition $\forall C^\sharp  \in
\mathcal{C}^\sharp$, $\mathbb{F}\circ\conc{}(C^\sharp)\subseteq
\conc{}\circ\counterpart{}(C^\sharp)$. 

Then, we extrapolate the iterates of $\counterpart{}$.
We define the abstract iteration \cite{cc:frameworks,cc:galois-widening} of $\counterpart{}$ as $\iteraten{}{0} \bydef \abot{}$  
 and $\iteraten{}{n+1} \bydef \iteraten{}{n} {\wid{}} \counterpart{}(\iteraten{}{n})$ for any $n\geq 0$.
The abstract iteration $(\iteraten{}{n})_{n\in\mathbb{N}}$
is ultimately stationary.
Moreover, its limit $\limit{}$ satisfies $\Coll\subseteq
\conc{}(\limit{})$ because $\mathbb{F}$ is monotonic.

\subsection{Coalesced product}
\label{product}
Several abstractions can be composed to refine each other.
We consider two abstractions:
\begin{equation*}
\begin{array}{l}
\abstraction{1}=\tuple{1}, \text{ and } \cr
\abstraction{2}=\tuple{2}.
\end{array}
\end{equation*}
We define the coalesced product between the abstractions $\abstraction{1}$ and $\abstraction{2}$ as the tuple $\tuple{}$, 
where the domain $\adom{}$ is defined as $\adom{1} \times \adom{2}$; 
the concretization $\conc{}$ is defined as the intersection of the two concretizations (i.e.~$\conc{}(a,b)\bydef\conc{1}(a)\cap\conc{2}(b)$);
the abstract union $\aunion{}$, the element $\abot{}$, the widening operator $\wid{}$, and the abstraction $\ainit{}$ of the initial state  are all defined pairwise; the abstract element $\abst{}((C_1,C_2),\overline{\lambda})$ is defined as 
$\bot$ whenever either $\abst{1}(C_1,\overline{\lambda})=\abot{1}$ or 
$\abst{2}(C_2,\overline{\lambda})=\abot{2}$, and as
$(\abst{1}(C_1,\overline{\lambda}),\abst{2}(C_2,\overline{\lambda}))$ otherwise.
The coalesced product between $\abstraction{1}$ and $\abstraction{2}$ is also an abstraction. 
We stress on the fact that the coalesced product is more powerful than a mere product. Thanks to the extended labeled transition system, several analyses can share 
constraints about the threads that are involved 
in computation steps. This way, analyses refine each other. 

We use our framework with the coalesced product between an analysis of the dynamic linkage between  threads (Sect.~\ref{environment}) and an analysis of each computation unit contents (Sect.~\ref{occurrence}).

\section{Environment analysis}
\label{environment}

\newcommand{\allconstraint}{\textit{Constraints}(\Vrec,\Veme)}
\newcommand{\constraintset}{C}
\newcommand{\constraint}{c}

We design an analysis of the dynamic linkage between threads. 
This analysis aims at capturing the relationship between the computation units of the threads that are involved in computation steps.

\subsection{Abstract domain}
Our goal is to map each program point label to an abstraction of the set of the marker/environment pairs which may be associated to any thread at this program point at run-time. 
So, we introduce for any set of variables $V\subseteq \Names$  a parametric abstract domain  $\Atome{V}$  of properties.
The concretization $\gammaatome{V}(a)$ of a property $a\in \Atome{V}$ is 
a set of marker/environment pairs m in $\wp(\env{V})$.
The operator $\cupatome{V}$ maps each finite set of properties to a weaker
property: for each finite set $A \subseteq \Atome{V}$, $\forall a\in A$, $\gammaatome{V}(a)\subseteq  \gammaatome{V}(\cupatome{V} A)$. 
The element $\botatome{V}$ is an abstraction of the empty set 
(i.e.~we assume that $\gammaatome{V}(\botatome{V})=\emptyset$).
The operator $\widatome{V}$ is a widening operator \cite{cc:galois-widening}.
Then, our main environment abstract domain $\Envdom$
is the set of the functions that map each program point label $l\in\prolabel$ that occurs in the system $\System$ to an element in $\Atome{\interface(l)}$. 
The domain structure ($\cupenv$, $\botenv$, and $\wid{\Envkey}$)  is defined point wise. 
The abstract domain $\Envdom$  is related to $\wp(\states)$ by
the concretization function $\gammaenv$ that maps each abstract property
$f\in\Envdom$ to the set of  configurations
$C\in\states$ such that $\forall (l,\textit{id},E)\in C,\;(\textit{id},E)\in \gamma_{\interface(l)}(f(l))$.

\begin{example}[\icfadomain]
\label{cfadomain}
We propose a simple \emph{cfa} domain to analyze the shared memory example (see Ex.~\ref{memory}).  In this example, the names that occur in computation units are never communicated. As a consequence, equality  among variables \cite[Sect.~5.1.1]{feret:esop2002}  and a uniform approximation of the control flow \cite{nielson:journal,nielson:concur98} are enough. In general, numerical abstractions of markers \cite{feret:sas2000,feret:esop2002}  are required. All these analyses \cite{nielson:journal,nielson:concur98,feret:esop2002,feret:sas2000} are polynomial time.

Given a set $\var$ of variables, we introduce the abstract domain 
$\flowdom{\var}$ as the set $\flowbot{\var} \uplus ((\var\mapsto \wp(\Labels))\times \wp(\var\times\{=;\not =\}\times \var))$. Each abstract element $e\in\flowdom{\var}$ denotes a set $\flowconc{\var}(e)\subseteq \var \mapsto \Labels\times \Markers$ of functions. More precisely, $\flowconc{\var}(\flowbot{\var})=\emptyset$ and $\flowconc{\var}(f,c)=\{g\;|\;
\forall (x,\diamond,y)\in c,\;g(x)\diamond g(y) \text{\; and\; } \forall x\in\var,\;\exists\textit{id}\in\Markers,\;g(x)=(f(x),\textit{id})\}$. This way, in the abstract element $(f,c)$, the function $f$ describes constraints about the label of values and the set $c$ describes constraints about equality and inequality relations among values.

We define a partial order $\flowsub{\var}$ over $\flowdom{\var}$ as: $\bot \flowsub{\var} x$, for any $x\in\flowdom{\var}$, and $(f_1,c_1) \flowsub{\var} (f_2,c_2)$ if and only if both $f_1(x)\subseteq f_2(x)$ and $c_2\subseteq c_1$. 
We notice that the concretization $\flowconc{\var}$ is monotonic with respect to $\flowsub{\var}$. Several abstract elements may have the same concretization, nevertheless, for each abstract element $e\in\flowdom{\var}$, the set of the elements $e'\in\flowdom{\var}$ such that $\flowconc{\var}(e)=\flowconc{\var}(e')$ has a least element that we denote $\rho(e)$. The element $\rho(e)$ is called the normal form of $e$. We denote by $\normalflowdom{\var}$ the set $\{\rho(e)\;|\;e\in\flowdom{\var}\}$ of all normal forms. We denote by $\normalflowsub{\var}$ the restriction of $\flowsub{\var}$ to $\normalflowdom{\var}$. Each subset $A\subseteq \normalflowdom{\var}$ has a least upper bound with respect to $\normalflowsub{\var}$, that we denote by $\normalflowcup{\var}$. 

The domain $\normalflowdom{V}$ is a good candidate for $\Atome{V}$. 
We also set $\gammaatome{V}\bydef[a \mapsto \Markers \times \normalflowconc{V}(a)]$, 
$\cupatome{V}\bydef\normalflowcup{V}$, and $\botatome{V}\bydef\flowbot{V}$. Since $\normalflowdom{V}$ is a finite domain, we define the widening operator $\widatome{V}$ as $a\widatome{V} b \bydef \normalflowcup{V}\{a;b\}$.
\boxexample\end{example}

\begin{figure*}[p]
\subfigure[Initial configuration abstraction.]
{\label{initenv}
\begin{minipage}{0.99\linewidth}
\begin{equation*}
\initenv\bydef \left[\begin{array}{l}
l\not\in\Extract(\System) \mapsto   \botatome{\interface(l)},\cr
l\in\Extract(\System) \mapsto 
  \declare_{x_n}(\ldots(\declare_{x_1}(\varepsilon_{\emptyset}))\ldots),   

(\text{where }  \{x_1,\ldots,x_n\}\bydef\interface(l))\end{array}\right].
\end{equation*}
\end{minipage}}
\subfigure[Abstract \textsc{post} operator.]{
\label{postenv}
\begin{minipage}{0.99\linewidth}
 Let $\lrec$ and $\leme$ be two program point labels in  $\prolabel$, such that
$\type(\lrec)\in\{\inputm,\fetch\}$, $\type(\leme)=\outputm$, and such that the length of the lists $\args(\lrec)$ and $\args(\leme)$ is the same.
We denote  $[y_1,\ldots,y_n]=\args(\lrec)$ and $[x_1,\ldots,x_n]=\args(\leme)$. 
Let $(\sim,A)\in\abstractcontext(\lrec,\leme)$ be a partition case and $\envf\in\Envdom$ be an abstract element.

We define:
\begin{itemize}
\item $\textsc{input}_0\bydef\envf(\lrec)$ and $\textsc{output}_0\bydef\envf(\leme)$;

\item $\textsc{input}_1 \bydef \begin{cases}
  \fetchm{\leme}{\textsc{input}_0} & \text{whenever } \type(\lrec)=\fetch,\cr
\textsc{input}_0 &\text{otherwise};
\end{cases}$
\item $\textsc{input}_3 \bydef \declare_{u_o}(\ldots(\declare_{u_1}(\extension_{y_n}(\ldots(\extension_{y_1}(\textsc{input}_1))\ldots)))\ldots)$

where  $\{u_1;\ldots;u_o\}\bydef (\bigcup \{\interface(l)\;|\;l\in \Extract(\cont(\lrec))\})\setminus \FN(\cont(\lrec))$,

\item $\textsc{output}_3\bydef \declare_{v_{p}}(\ldots(\declare_{v_1}(\textsc{output}_0))\ldots)$,

where $\{v_1;\ldots;v_{p}\}\bydef(\bigcup \{\interface(l)\;|\;l\in\Extract(\cont(\leme))\}) \setminus \FN(\cont(\leme))$;

\item $\textit{mol}_0 \bydef \textsc{input}_3\concat \textsc{output}_3$;

\item $\textit{cons} \bydef \textit{com} \cup \textit{part$_=$} \cup \textit{part$_{\not =}$} \cup \textit{part$_\text{lbl}$}$, 
where:

$\textit{com} \bydef
\{(\canal(\lrec),\rec)=(\canal(\leme),\eme)\}
\cup\{(y_k,\rec)=(x_k,\eme)\;|\;1\leq k \leq n\}$,

$\textit{part$_=$} \bydef \{(\giveindex(l_1)(b),\diamond_1)=(\giveindex(l_2)(b),\diamond_2)\;|\;(l_1,\diamond_1),(l_2,\diamond_2)\in \vartrans,\;b\in B,\;(l_1,\diamond_1)\sim (l_2,\diamond_2)\}$,

$\textit{part$_{\not =}$} \bydef \{(\giveindex(l_1)(b),\diamond_1)=(\giveindex(l_2)(b),\diamond_2)\;|\;(l_1,\diamond_1),(l_2,\diamond_2)\in \vartrans,\;(l_1,\diamond_1)\not\sim (l_2,\diamond_2),\;B_s=\{b\}\}$,

$\textit{part$_\text{lbl}$} \bydef \{\textrm{lbl}((\giveindex(l)(b),\diamond),A([(l,\diamond)]_{\sim})(b))\;|\;(l,\diamond)\in \vartrans\}$;
\item  $\textit{mol}_1 \bydef \sync{\textit{cons}}{\textit{mol}_0}$;

\item $\abst{\Envkey}(\envf,((\lrec,\leme),(\sim,A))) \bydef
\begin{cases} 
\botenv & \text{if  } \textit{mol}_1=\botmolecule{\Vrec}{\Veme},\cr
\cupenv \{\envf;\envf'\}&\text{otherwise}, 
\end{cases}$

where $\envf' \bydef \begin{cases}
l \mapsto \gc{\interface(l)}{\fst(\textit{mol}_1)} & \text{whenever } l\in\Extract(\cont(\lrec)),\cr
l \mapsto \gc{\interface(l)}{\snd(\textit{mol}_1)} & \text{whenever } l\in\Extract(\cont(\leme)),\cr
l \mapsto \bot_{\interface(l)} & \text{otherwise}.\end{cases}$
\end{itemize}
\end{minipage}}
\caption{Environment analysis.}
\label{abstractenv}
\end{figure*}

Now, we simulate the non-standard semantics in the abstract.
\subsection{Initial state}
At the beginning of the concrete computation, 
the configuration contains one thread at each program point the label of which is in the set $\Extract(\System)$. Thread markers are $\varepsilon$ and environments map each free variable $x$ to the name $(x,\varepsilon)$. In the abstract, we require two primitives.
First, the abstract property $\varepsilon_{\emptyset}\in\Atome{\emptyset}$ is the abstraction of the pair $(\varepsilon,\emptyset)$. This means that:  $\{(\varepsilon,\emptyset)\}\subseteq\gammaatome{\emptyset}(\varepsilon_\emptyset)$.
Then, the  primitive $\jdeclare$ simulates name allocation. 
Let $V$ be a set of variables and $x\in\Names\setminus V$ be a fresh variable. 
The primitive $\declare_x$ is a function in $\Atome{V}\rightarrow \Atome{V\cup\{x\}}$ and, for any abstract element $a\in \Atome{V}$, 
the concretization $\gammaatome{V\cup \{x\}}(\declare_x(a))$ contains at least all pairs  $(\textit{id},E)\in \env{V\cup \{x\}}$ such that (i) $(\textit{id},E_{|V})\in\gamma_V(a)$, (ii) $E(x)=(x,\textit{id})$, and (iii) $\forall y\in V, E(y)\not = E(x)$.

\begin{example}[\ccfadomain]
In our simple \emph{cfa} domain (see Ex.~\ref{cfadomain}), 
the primitive $\varepsilon_{\emptyset}$ can be defined as $(\emptyfun,\emptyset)$ (where, in the first component, the symbol $\emptyfun$ denotes the function defined over the empty set). Moreover, we define $\declare_x$ by: 
$\declare_x(\botatome{V})\bydef\botatome{V}$  and by $\declare_x((f,c))\bydef\rho(f',c')$ where $f'\bydef f[x\mapsto x]$ and $c'\bydef c\cup \{(x,\not=,a)\;|\;a\in\ V\}$. This means that we know that the channel has been opened by an instance of a restriction $(\nuu x)$ and we know that this value is fresh. Then, we apply our closure $\rho$.
\boxexample\end{example}

The abstraction $\initenv\in\Envdom$ of initial state is defined in Fig.~\ref{initenv} as the function that maps any program point $l$ to the abstract element $\declare_{x_n}(\ldots(\declare_{x_1}(\varepsilon_{\emptyset}))\ldots)$ whenever 
$l\in\Extract(\System)$ and $\{x_1;\ldots;x_n\}\bydef \interface(l)$; 
and to the abstract element $\botatome{\interface(l)}$ otherwise.

\begin{example}[\cflowanalysis]
We apply our analysis with the simple \emph{cfa} abstract domain (e.g.~see Ex.~\ref{cfadomain}) on the shared memory system (e.g.~see Ex.~\ref{memory}).
We obtain that: $\initenv(\memlaba)=\rho([\ccreate \mapsto \ccreate,\cnull \mapsto \cnull],\emptyset)$, $\initenv(\memlabl)=\rho([\crec_{\memlabl}\mapsto \crec_{\memlabl}],\emptyset)$, $\initenv(\memlabl')=\rho([\ccreate\mapsto \ccreate,\crec_{\memlabl}\mapsto \crec_{\memlabl}],\emptyset)$, and $\initenv(l)=\botatome{\interface(l)}$ 
for any $l\not\in\{\memlaba;\memlabl;\memlabl'\}$.
\boxexample\end{example}

\subsection{Transition step}
In the concrete, an interaction involves two threads: $\trec$ at a program point labeled with  $\lrec$  and $\teme$ at a program point labeled with $\leme$. 
The first thread either inputs a message or fetches a resource; 
the second thread outputs a message.
We simulate such a transition $\tau$ in the abstract in Fig.~\ref{postenv}. 
We start from the abstract element $\envf\in\Envdom$ and we define the pair $(\sim,A)\in\abstractcontext(\lrec,\leme)$ as $\alpha_{\textsc{step}}(\tau)$.

\begin{example}[\cflowanalysis]
We apply our analysis with the simple \emph{cfa} abstract domain (e.g.~see Ex.~\ref{cfadomain}) on the shared memory system (e.g.~see Ex.~\ref{memory}).
As an example, we focus on the interaction between a thread at program point \pp{\memlabe} and a thread at  program point \pp{\memlabj} in any calling context $(\sim,A)\in\abstractcontext(\lrec,\leme)$.
We also assume that the element 
$\envf(\memlabe)$ is equal to $\rho([\ccell \mapsto \{\ccell\}, \port\mapsto\{\addread\}],\emptyset)$ and 
that the  element $\envf(\memlabj)$ is equal to $\rho([\ccell \mapsto \{\ccell\}, \datawrite \mapsto \{\cdata\}],\emptyset)$.

We want to prove that:
\begin{itemize}
\item  both that interact and the thread that is launched at program point \pp{\memlabf} belong to the same partition class (i.e.~$(\memlabe,\rec)\sim (\memlabf,\rec)$, $(\memlabe,\rec)\sim(\memlabj,\eme)$);
\item the thread that interacts at the program point \pp{\memlabe} and the thread that is launched at the program point \pp{\memlabg} do not belong to the same partition class (i.e.~$(\memlabe,\rec)\not\sim(\memlabg,\rec)$);
\item  and that the abstraction of the computation unit of interacting threads is $[b\mapsto \ccell]$ (i.e.~$A([(\memlabe,\rec)]_\sim)(b)=\ccell$). 
\end{itemize}
Then, we want to abstract the environment of the thread that is launched at program point \pp{\memlabf}.
\boxexample\end{example}

\subsubsection{Extending environments}
First, we collect information about the potential binding of the  threads $t_\rec$ and $t_\eme$. We denote by $\textsc{input}_0$ the element $\envf(\lrec)$ and by $\textsc{output}_0$ the element $\envf(\leme)$. 
In the concrete, a new thread marker is computed when the input thread is a resource. 
We require a primitive $\jfetchm$ to simulate the allocation of this fresh marker in the abstract. 
For any set $V\subseteq \Names$ of variables, any abstract element $a\in\Atome{V}$, and any label $l\in\Labels$, the abstract element $\fetchm{l}{a}\in\Atome{V}$ satisfies: the concretization $\gammaatome{V}(\fetchm{l}{a})$ 
contains at least all pairs $(l.\textit{id},E)\in \env{V}$ such that 
 $(\textit{id},E)\in\gammaatome{V}(a)$.

\begin{example}[\ccfadomain]
In our simple \emph{cfa} domain (see Ex.~\ref{cfadomain}), 
we do not track any information about thread markers.
So we define the element $\jfetchm(l,a)$ as $a$.
\boxexample\end{example}

Then, we define $\textsc{input}_1$ as $\fetchm{\leme}{\textsc{input}_0}$ whenever the thread $\trec$ is a resource (i.e.~if $\type(\lrec)=\fetch$), and 
 as $\textsc{input}_0$ otherwise.

\begin{example}[\cflowanalysis]
In our example, we have:
\begin{equation*}
\begin{array}{l}
 \textsc{input}_1 = \rho([\ccell  \mapsto \{\ccell\},\;\port  \mapsto\{\addread\}],\emptyset) \text{ and }\cr 
\textsc{output}_0 = \rho([\ccell \mapsto \{\ccell\}, \datawrite \mapsto \{\cdata\}],\emptyset).
\end{array}
\end{equation*}
\boxexample\end{example}

We now extend the environments to deal with the variables introduced during the interaction.
In the concrete, the threads $\trec$ and $\teme$  bind some new variables to some names. 
The sequence $[y_1,\ldots,y_n]\bydef\args(\lrec)$ is the sequence of the variables that are bound by name passing.
We use an abstract primitive $\extension$ to create these variables without any information about them. 
For any set $V\subseteq \Names$ of variables, any variable $x\not\in V$, and any abstract element $a\in\Atome{V}$, the abstract element $\extension_x(a)\in\Atome{V\cup \{x\}}$ satisfies: $\{(\textit{id},E)\in \env{V\cup \{x\}}\;|\;(\textit{id},E_{|V})\in\gammaatome{V}(a)\} \subseteq \gammaatome{V\cup \{x\}}(\extension_x(a))$. 

\begin{example}[\ccfadomain]
We can define the primitive $\extension$ by $\extension_x(\botatome{V})\bydef\botatome{V}$ and by $\extension_x(f,c)\bydef(f[x\mapsto \Labels],c)$. 
\boxexample\end{example}

Thus,  we define $\textsc{input}_2$ by $\extension_{y_n}(\ldots(\extension_{y_1}(\textsc{input}_1))\ldots)$. The set of the variables that are 
bound by name restriction in the thread $\trec$ is given by $\{u_1;\ldots;u_o\}\bydef (\bigcup \{\interface(l)\;|\;l\in \Extract(\cont(\lrec))\})\setminus \FN(\cont(\lrec))$, whereas the one in the thread $\teme$ is given by $\{v_1;\ldots;v_{p}\}\bydef(\bigcup \{\interface(l)\;|\;l\in\Extract(\cont(\leme))\}) \setminus \FN(\cont(\leme))$. We introduce these variables thanks to the primitive $\jdeclare$. We define $\textsc{input}_3 \bydef \declare_{u_o}(\ldots(\declare_{u_1}(\textsc{input}_2))\ldots)$ and  $\textsc{output}_3\bydef \declare_{v_{p}}(\ldots(\declare_{v_1}(\textsc{output}_0))\ldots)$.

\begin{example}[\cflowanalysis]
In the shared memory example, the variable $\dataread$ is bound during the communication. 
Moreover, since  no variable is bound by a name restriction, the abstract element 
$\textsc{input}_3$ is equal to $\rho(f,\emptyset)$  
where $f=[\ccell  \mapsto  \{\ccell\},\port  \mapsto \{\addread\},\dataread  \mapsto  \Labels]$, and the abstract element 
$\textsc{output}_3$ is equal to $\rho([\ccell \mapsto \{\ccell\}, \datawrite \mapsto \{\cdata\}],\emptyset)$.
\boxexample\end{example}

To get precise relations between the binding of former variables and the binding of the variables bound by the communication, we gather the two descriptions $\textsc{input}_3$ and $\textsc{output}_3$. 
For that purpose, we assume that we are given, for any subset of variables $\Vrec$,$\Veme\subseteq \Names$, 
an abstract domain $\Molecule{\Vrec}{\Veme}$ of properties about 
sets of pairs of marker/environment pairs.
Each property in $\Molecule{\Vrec}{\Veme}$ is related by a concretization function $\gammamolecule{\Vrec}{\Veme}$ to the elements of $\wp(\env{\Vrec}\times\env{\Veme})$ which satisfy this property.
We also introduce an element $\botmolecule{\Vrec}{\Veme}$ that satisfies $\gammamolecule{\Vrec}{\Veme}(\botmolecule{\Vrec}{\Veme})=\emptyset$.
The domains $\Atome{\Vrec}$, $\Atome{\Veme}$, and $\Molecule{\Vrec}{\Veme}$ are related by the following primitives. 
The primitive $\concat$ simulates pair construction.
For any $a_\rec\in \Atome{\Vrec}$ and any $a_\eme\in\Atome{\Veme}$, the element $a_\rec\concat a_\eme \in\Molecule{\Vrec}{\Veme}$ satisfies: $\gammaatome{\Vrec}(a_?)\times\gammaatome{\Veme}(a_!)\subseteq \gammamolecule{\Vrec}{\Veme}(a_?\concat a_!)$;
the primitives $\fst$ and $\snd$ abstract the projection functions: 
for any $a\in\Molecule{\Vrec}{\Veme}$, the elements $\fst(a)\in\Atome{\Vrec}$ and $\snd(a)\in\Atome{\Veme}$ satisfy:
 $\gammamolecule{\Vrec}{\Veme}(a) \subseteq \!\gammaatome{\Vrec}(\fst(a))\times\gammaatome{\Veme}(\snd(a))$.

Then, we gather the two properties thanks to the abstract product $\concat$. 
We define $\textit{mol}_0$ as $\textsc{input}_3\concat \textsc{output}_3$. 
We denote by $(\Vrec,\Veme)\in\wp(\Names)^2$ the pair of sets of variables such that $\textit{mol}_0 \in\Molecule{\Vrec}{\Veme}$. The element $\textit{mol}_0$ abstracts a set of pairs $((\textit{id}_\rec,\Erec),(\textit{id}_\eme,\Eeme))\in \env{\Vrec}\times \env{\Veme}$. We introduce some formal variable to denote the channel names that are bound  either in the environment $\Erec$, or in the environment $\Eeme$. We introduce the set $\textit{Var}(\Vrec,\Veme)\bydef\{(v,?)\;|\;v\in \Vrec\}\cup \{(v,!)\;|\;v\in \Veme\}$ of formal variables. 

\begin{example}[\ccfadomain]
We can define the abstract domain $\Molecule{\Vrec}{\Veme}$ as $\normalflowdom{\textit{Var}(\Vrec,\Veme)}$. 
The concretization $\gammamolecule{\Vrec}{\Veme}$ maps each abstract element $a$ to the set of pairs $(\textit{id}_\rec,\Erec),(\textit{id}_\eme,\Eeme)$ such that the map $[(x,\rec)\mapsto \Erec(x),(x,\eme)\mapsto \Eeme(x)]$ belongs to $\normalflowconc{\textit{Var}(\Vrec,\Veme)}(a)$. 
The bottom element $\botmolecule{\Vrec}{\Veme}$ can be defined as $\flowbot{\textit{Var}(\Vrec,\Veme)}$.

The primitive $\fst$ maps 
$\botmolecule{\Vrec}{\Veme}$ to $\botatome{\Vrec}$ and any other element $(f,c)$ to the element $([x\in\Vrec\mapsto f(x,\rec)],\{(x,\diamond,y)\;|\;((x,\rec),\diamond,(y,\rec))\in c\})$. 
The primitive $\snd$  maps $\botmolecule{\Vrec}{\Veme}$ to $\botatome{\Veme}$
 and any  other element $(f,c)$ to the element $([x\in\Veme\mapsto f(x,\eme)],\{(x,\diamond,y)\;|\;((x,\eme),\diamond,(y,\eme))\in c\})$. 
The abstract product is defined by: 
$\botatome{\Vrec}\concat e_\eme = e_\rec \concat \botatome{\Veme} = \botmolecule{\Vrec}{\Veme}$ and by 
$(f_\rec,c_\rec) \concat (f_\eme,c_\eme) \bydef (f',c')$, where $f'\bydef[(x,i) \mapsto f_i(x)]$ and $c'\bydef\{((x,i),\diamond,(y,i))\;|\;(x,\diamond,y)\in c_i\}$.
\boxexample\end{example}

\begin{example}[\cflowanalysis]
In our example, the abstract element $\textit{mol}_0$ is equal to $\rho(f,\emptyset)$ where the function  $f$ is defined as the following function:
\begin{equation*}
\begin{cases}
(\ccell,\rec)  \mapsto \{\ccell\},\cr 
(\port,\rec) \mapsto\{\addread\},\cr
(\dataread,\rec) \mapsto \Labels,\cr
(\ccell,\eme) \mapsto \{\ccell\},\cr
(\datawrite,\eme) \mapsto \{\cdata\}.\cr
\end{cases}
\end{equation*}
\boxexample\end{example}

\subsubsection{Collecting new constraints}

Now, we collect the set $\textit{cons}$ of all the constraints that we have about the environments $\Erec$ and $\Eeme$.
The formal variable  $(v,?)$ denotes the value $\sigma(v,?)\bydef\Erec(v)$ of the variable $v$ in the input thread and the variable $(v,!)$ denotes the value $\sigma(v,!)\bydef\Eeme(v)$ of the variable $v$ in the output thread.
We consider three kinds of constraints:
the constraint $v_1 = v_2$  where $v_1,v_2\in\textit{Var}(\Vrec,\Veme)$ means that the formal variables $v_1$ and $v_2$ denote the same channel name: we write $p\sat v_1 = v_2$ if and only if $\sigma(v_1)=\sigma(v_2)$;
the constraint $v_1 \not= v_2$  is the negation of the constraint $v_1=v_2$: we write $p\sat v_1 \not= v_2$ if and only if $\sigma(v_1)\not =\sigma(v_2)$;
the constraint $\textrm{lbl}(v,l)$, where $v\in\textit{Var}(\Vrec,\Veme)$ and $l\in\Labels$ means that $l$ is the label of the name that is denoted by the formal variable $v$: 
we write $p\sat \textrm{lbl}(v,l)$ if and only if 
$\sigma(v)$ matches $(l,\_)$.
We denote by $\allconstraint$ the set of all such constraints.
First, we collect the constraints due to communication: 
the constraint $(\canal(\lrec),\rec)=(\canal(\leme),\eme)$ encodes the fact that both threads interact over the same channel and 
the set $\{(y_k,\rec)=(x_k,\eme)\;|\;1\leq k \leq n\}$ of constraints encodes name-passing. 
Now we consider the constraints given by $\sim$:
for any pair $(l_1,\diamond_1),(l_2,\diamond_2)\in \vartrans$ such that $(l_1,\diamond_1)\sim (l_2,\diamond_2)$,  
the set of constraints  $\{(\giveindex(l_1)(b),\diamond_1)=(\giveindex(l_2)(b),\diamond_2)\;|\;b\in B\}$ encodes the fact that the threads that are denoted by the pairs $(l_1,\diamond_1)$ and $(l_2,\diamond_2)$ share the same computation unit; 
conversely, when $B_s$ is not a singleton, we cannot extract constraints from non-equality among computation units, but  
when $B_s$ is a singleton $\{b\}$, for any pairs $(l_1,\diamond_1),(l_2,\diamond_2)\in \vartrans$ such that $(l_1,\diamond_1)\not \sim (l_2,\diamond_2)$, the constraint $(\giveindex(l_1)(b),\diamond_1)\not =(\giveindex(l_2)(b),\diamond_2)$ encodes the fact that the threads that are denoted by the pairs $(l_1,\diamond_1)$ and $(l_2,\diamond_2)$ are not in the same computation unit; 
last, for any pair $(l,\diamond)\in \vartrans$, the set of constraints $\{\textrm{lbl}((\giveindex(l)(b),\diamond),A([(l,\diamond)]_{\sim})(b))\;|\;b\in B\}$ models the fact that $A([(l,\diamond)]_{\sim})$ 
is the abstract computation unit of the thread denoted by the pair $(l,\diamond)$.

\begin{example}[\cflowanalysis]
In our example, we get the constraint set $\textit{com}\cup\textit{part}_=\cup\textit{part}_{\not =}\cup \textit{part}_{\text{lbl}}$, where:
\begin{equation*}
\textit{com}=\{(\ccell,\rec)=(\ccell,\eme);(\dataread,\rec)=(\datawrite,\eme)\},
\end{equation*} and $\textit{part}_=$, $\textit{part}_{\not =}$, and $\textit{part}_{\text{lbl}}$ are defined as in Fig.~\ref{postenv} (they depend on the pair $(\sim,A)$).
\boxexample\end{example}

We can now define $\textit{mol}_1$ as $\sync{\textit{cons}}{\textit{mol}_0}$, where the primitive $\jsync$ is used to enforce some constraints in abstract elements. For any set $\constraintset\in \allconstraint$ of constraints and any abstract element $a\in\Molecule{\Vrec}{\Veme}$, the element $\sync{\constraintset}{a}\in \Molecule{\Vrec}{\Veme}$ 
is such that the set  $\gammamolecule{\Vrec}{\Veme}(\sync{\constraintset}{a})$ contains at least all pairs $p=((\idrec,\Erec),(\ideme,\Eeme))$ that satisfy both  $p \in \gamma_{(\Vrec,\Veme)}(a)$ and  $\forall \constraint \in \constraintset, p \sat \constraint$. 

\begin{example}[\ccfadomain]
We can define the primitive $\jsync$ as follows:
\begin{equation*}
\begin{cases}
\sync{\textit{cons}}{\botmolecule{\Vrec}{\Veme}}\bydef{\botmolecule{\Vrec}{\Veme}},\cr
\sync{\textit{cons}}{(f,c)}\bydef\rho(f',c\cup\{(x,\diamond,y)\;|\;x\diamond y\in \textit{cons}\}),\cr
\end{cases}
\end{equation*}
where
$
f'=\begin{cases}
x\mapsto f(x) & \text{whenever } \nexists l,\; \textrm{lbl}(v,l)\in\textit{cons},\cr
x\mapsto \{l\} & \text{whenever } !\exists l,\; \textrm{lbl}(v,l)\in\textit{cons},\cr
x\mapsto \emptyset & \text{otherwise};
\end{cases}$

\noindent We stress that the normalization step is crucial to propagate information, and especially to detect unsatisfiable constraints.
\boxexample\end{example}

\begin{example}[\cflowanalysis]
First, we prove that the interaction is not possible as soon as 
$(\memlabe,\rec)\not \sim (\memlabf,\rec)$, $(\memlabe,\rec)\not \sim(\memlabj,\eme)$, $(\memlabe,\rec)\sim(\memlabg,\rec)$, or   $A([(\memlabe,\rec)]_\sim)(b)\not =\ccell$:
\begin{itemize} 
\item If $(\memlabe,\rec)\not \sim (\memlabf,\rec)$, we have  $(\ccell,\rec)\not=(\ccell,\rec)\in\textit{part}_{\not =}$. Then,  $\textit{mol}_1 = \botmolecule{\Vrec}{\Veme}$.
\item If $(\memlabe,\rec)\not \sim (\memlabj,\eme)$, we have  $(\ccell,\rec)\not=(\ccell,\eme)\in\textit{part}_{\not =}$. 
But $(\ccell,\rec)=(\ccell,\eme)\in\textit{com}$. 
Then, $\textit{mol}_1 =\botmolecule{\Vrec}{\Veme}$.
\item 
If $(\memlabe,\rec)\sim(\memlabg,\rec)$, we have $(\ccell,\rec)\sim(\port,\rec)\in\textit{part}_=$.
Then, $\textit{mol}_1$ matches $\rho(f,c)$ with $(\ccell,\rec)\sim(\port,\rec)\in c$, $f(\ccell,\rec)=\{\ccell\}$, and $f(\port,\rec)=\{\addread\}$.
Since $f(\ccell,\rec)\cap f(\port,\rec)=\emptyset$, we have $\textit{mol}_1 = \botmolecule{\Vrec}{\Veme}$.
\item 
If  $A([(\memlabe,\rec)]_\sim)(b)\not =\ccell$, 
 $\{\textrm{lbl}((\ccell,\rec),A([(\memlabe,\rec)]_{\sim})(b)\}\in\textit{part}_{\text{lbl}}$. Then $\textit{mol}_1$ matches $\rho(f,c)$ with 
$f(\ccell,\rec)=\{\ccell\}\cap\{A([(\memlabe,\rec)]_{\sim})(b)\}$. So 
$\textit{mol}_1=\botmolecule{\Vrec}{\Veme}$.
\end{itemize}

Until the end of the section, we assume that: $(\memlabe,\rec) \sim (\memlabf,\rec)$, $(\memlabe,\rec) \sim (\memlabj,\eme)$, $(\memlabe,\rec)\not\sim (\memlabg,\rec)$, and $A([(\memlabe,\rec)]_\sim)(b)=\ccell$. 

With these assumptions, we have:
\begin{itemize}
\item $\textit{com}=\{(\ccell,\rec)=(\ccell,\eme);(\dataread,\rec)=(\datawrite,\eme)\}$, 
\item $\textit{part}_= = \{(\ccell,\rec)=(\ccell,\eme);(\ccell,\rec)=(\ccell,\rec)\}$, 
\item $\textit{part}_{\not =}=\{(\ccell,\rec)\not = (\port,\rec)\}$, 
\item $\textit{part}_{\text{lbl}}=\left\{\begin{array}{l}\textrm{lbl}(\ccell,\rec),\ccell);\textrm{lbl}((\ccell,\eme),\ccell);\cr
\textrm{lbl}((\port,\rec),A([(\memlabg,\rec)]_\sim))\end{array}\right\}$.
\end{itemize}
Then, $\textit{mol}_1 = \rho(f,\textit{com}\cup\textit{part}_=\cup \textit{part}_{\not =})$ 
where the function $f$ is defined as 
\begin{equation*}
\begin{cases}
(\ccell,\rec) \mapsto \{\ccell\},\cr 
(\port,\rec) \mapsto\{\addread\},\cr
(\dataread,\rec) \mapsto \Labels,\cr
(\ccell,\eme) \mapsto \{\ccell\},\cr
(\datawrite,\eme) \mapsto \{\cdata\}.
\end{cases}
\end{equation*}
Since the constraint $(\datawrite,\eme)=(\dataread,\rec)$ belongs to 
the set $\textit{com}$ of constraints, we can deduce that the abstract element $\textit{mol}_1$ is equal to $\rho(f',\textit{com}\cup\textit{part}_=\cup \textit{part}_{\not =})$, where $f'=f[(\dataread,\rec) \mapsto \{\cdata\}]$.
\boxexample\end{example}

\subsubsection{Updating the abstract element}

Whenever we have $\textit{mol}_1=\botmolecule{\Vrec}{\Veme}$, the constraints are not satisfiable, so we set $\abst{\Envkey}(\envf,((\lrec,\leme),(\sim,A))) \bydef\botenv$. Otherwise, we first separate information about the input and the output threads, then we update the information about the threads that are launched. For that purpose, we use a primitive  $\jgc$ to simulate garbage collection: for any sets $X,V$ of variables such that $X\subseteq V$, and any abstract element $a\in\Atome{V}$, the  abstract element $\jgc_X(a)\in\Atome{X}$ satisfies the property $\{(\textit{id},E_{|X})\in\env{X}\;|\;(\textit{id},E)\in\gammaatome{V}(a)\}\subseteq \gammaatome{X}(\jgc_X(a))$.

\begin{example}[\ccfadomain]
The primitive $\jgc$ can be defined by  $\jgc_X(\botatome{V})\bydef\botatome{X}$ and  by $\jgc_X(f,c)\bydef(f_{|X},c\cap X\times\{=;\not=\}\times X)$.
\boxexample\end{example}

We define the element $\abst{\Envkey}(\envf,((\lrec,\leme),(\sim,A)))$  by $\cupenv \{\envf;\envf'\}$, where $\envf'(l)\bydef \gc{\interface(l)}{\fst(\textit{mol}_1)}$ whenever the label $l$ belongs to  the set $\Extract(\cont(\lrec))$, 
 $\envf'(l)\bydef \gc{\interface(l)}{\snd(\textit{mol}_1)}$ whenever 
the label $l$ is in the set $\Extract(\cont(\leme))$, 
and  $\envf'(l)\bydef\bot_{\interface(l)}$ otherwise.

\begin{example}[\cflowanalysis]
In our example, the function $\envf'$ satisfies: $\envf'(\memlabf)$  is equal to the element $\rho(f,\emptyset)$, where $f= [(\ccell,\rec) \mapsto \{\ccell\},\;(\dataread,\rec) \mapsto \{\datawrite\}]$. This is a precise abstraction of the environment of the thread that is launched at the program point \pp{\memlabf}.
\end{example}

\subsection{Soundness}

Thm.~\ref{envsound} states the soundness of our environment analysis.
\begin{theorem}
\label{envsound}
$\tuple{\Envkey}$ is an abstraction.
\end{theorem}

\section{Contents  analysis}
\label{occurrence}

Contents analysis counts both the number of threads inside each computation unit and the number of computation steps in the history of computation units. 
Its main goal is to detect mutual exclusion of threads inside computation units.

\subsection{Abstract domain}
Let $\var$ be the  set of variables $\{x_l\;|\;l\in \prolabel\}\cup
\{y_\lambda\;|\;\lambda\in \prolabel^2\}\cup \{z_\lambda\;|\;\lambda\in\prolabel^2\}$.
We use these variables to abstract both the contents and the history of computation units.
Given a computation unit: the variable $x_l$ counts the occurrence number of threads at the program point  $l$ in this computation unit,
 the variable $y_\lambda$ counts the number of computation steps labeled with $\lambda$ that have modified this computation unit, and the variable $z_\lambda$ is equal to $1$ if at least one computation step labeled with $\lambda$ has modified the contents of this computation unit and  is equal to $0$ otherwise.

We assume that we are given an abstract domain $\abstnum$ to abstract functions in $\var\rightarrow \mathbb{N}$.
Each abstract property is related to the set $\wp(\var\rightarrow \mathbb{N})$ by a  concretization $\gammanum$.
An operator $\cupnum$ maps each finite set of properties to a weaker
property: for each finite set $A \subseteq \wp(\abstnum)$,
$\forall a\in A$, $\gammanum(a)\subseteq \gammanum(\cupnum A)$. 
The element $\botnum$ is the abstraction of the empty set (i.e.~we have $\gammanum(\botnum)=\emptyset$).  The operator $\widnum$ is a widening \cite{cc:galois-widening}.
Then, our main  abstract domain 
$\Numdom$ is the set $\abstractunit \rightarrow \abstnum$ of the functions mapping each abstract computation unit to an abstraction of its contents.
The structure ($\cupNum$, $\botNum$, and $\wid{\Numkey}$)  is defined point wise. 
We define the concretization $\gammaNum(\tp)$ of any abstract element $\tp\in\Numdom$ as the set of  all configurations $C\in\states$ such that for any concrete computation unit $u\in\computationunit$, $\tp(\piunit(u))$ is an approximation of the contents of $u$. More precisely, we require that there exists  a map  $n\in\concnum(\tp(\piunit(u)))$ such that $\forall l\in\prolabel$, the number of  threads in $C$ at the program point  $l$ in the computation unit $u$ is equal to $n(x_l)$. We also require that, for any $\lambda\in\prolabel^2$, we have $n(z_{\lambda})=1$ whenever $n(y_{\lambda})\geq 1$, and  $n(z_{\lambda})=0$ otherwise (we require no further properties about the variables $y_\lambda$ and $z_{\lambda}$).

\begin{example}[\icount]
\label{intequ}
We propose to use a reduced product between the interval domain \cite{cc76} and the affine equality domain \cite{karr}. This way, our abstract domain expresses constraints either of the form $a\leq v\leq b$, or of the form $\sum  a_k.v_k = b$. Interval constraints (of the form $a\leq v\leq b$ where $a,b\in\mathbb{N}$ and $v\in\var$) express properties of interest. Affine equalities (of the form $\sum a_k.v_k = b$ where $a_1,\ldots,a_n,b\in\mathbb{Q}$, and $v_1,\ldots,v_n \in \var$ express more complex properties, such as mutual exclusion. This allows for more precise calculations in the interval domain. Moreover, affine equalities capture relations when some threads are created and some others are consumed. To get a good precision, we need  to avoid undetermined forms (when two unbounded values are subtracted) as much as possible.  So, we use the approximate reduced product given in \cite[Chap.~9]{feret:thesis}, in which each primitive can be computed in $\mathcal{O}(\textit{Card}(\var)^3)$ operations. Thus, we get a polynomial analysis. 

Other domains could have been considered. 
The polyhedron domain  \cite{ch} is too  expensive.
The octagon domain   \cite{mine:thesis,mine} cannot express the affine invariants that are required when dealing with semaphores that both involve more than two agents and several tokens.
Abstract multi-sets \cite{HJNN99ai,NiNi00popl} are exponential  in time.
\boxexample\end{example}

\begin{example}[\icontentanalysis]
We apply our content analysis on the example of the shared-memory (e.g.~see example \ref{memory}) with the reduced product of intervals and affine equalities (e.g.~see example \ref{intequ}).
We denote by $\tp$ the result of our analysis. 
The constraint system  $\tp(\ccell)$ describes the usage of channels opened by the instances of the restriction $\nuu \ccell$.
Our goal is to prove that the system $\tp(\ccell)$ entails both the affine equality constraint $x_{\memlabb} + x_{\memlabf} + x_{\memlabj} = y_{\memlaba,\memlabm}$ and the interval constraint $0\leq y_{\memlaba,\memlabm} \leq 1$.
This means that either the channel has not been opened yet (i.e.~$y_{\memlaba,\memlabm}=0$), or the channel has been opened (i.e.~$y_{\memlaba,\memlabm}=1$) and there is exactly one output over it at the program point \pp{\memlabb}, \pp{\memlabf}, or \pp{\memlabj} (since $x_{\memlabb} + x_{\memlabf} + x_{\memlabj}=1$).
\boxexample\end{example}

Now, we simulate the non-standard semantics in the abstract.

\subsection{Initial state}
\begin{figure*}[p]
\subfigure[Initial configuration abstraction.]{
\label{initocc}
\begin{minipage}{0.99\linewidth}
\begin{equation*}
\initnum \bydef [a \mapsto \cupnum\{\xinum(\{x_l\;|\;l\in \Extract(\System)\;|\;\giveindex(l)=a\}),\xinum(\emptyset)\}].
\end{equation*}
\end{minipage}}
\subfigure[Abstract \textsc{post} operator.]{
\label{postocc}
\begin{minipage}{0.996\linewidth}
Let $\lrec$ and $\leme$ be two program point labels in  $\prolabel$, such that
$\type(\lrec)\in\{\inputm,\fetch\}$, $\type(\leme)=\outputm$, and such that the length of the lists $\args(\lrec)$ and $\args(\leme)$ is the same.
Let $(\sim,A)\in\abstractcontext(\lrec,\leme)$ be a partition case and $\tp\in\domnum$ be an abstract element.
We define $\abst{\numkey}(\tp,((\lrec,\leme),(\sim,A)))$ by $\botNum$, whenever there exists $\diamond\in \{\lrec;\leme\}$ such that 
$\syncnume([(l_{\diamond},\diamond)]_{\sim}\cap \{(\lrec,\rec);(\leme,\eme)\})(\tp(A([(l_{\diamond},\diamond)]_\sim)))= \botnum$; otherwise, we define it by 
$[ a\mapsto \cupnum \{\tp(a)\}\cup \{\textit{content}_1(P)\;|\;P\in\classes,\;A(P)=a\}]$, where, for any $P\in \classes$:
\begin{itemize}
\item $\textit{old}(P) \bydef \xinum(\emptyset)$, 

$\text{ whenever } \{\giveindex(l)(b)\in \interface(l)\setminus \FN(\cont(l_\diamond))\;|\;(l,\diamond)\in P,\; b\in B\}\not = \emptyset, \text{ or }$ 
\item  $\syncnume(P\cap \{(\lrec,\rec);(\leme,\eme)\})(\tp(A(P))),  \text{ otherwise};$

\item $\textit{consumed}_{\rec}(P) \bydef \begin{cases} \{l_\rec\} & \text{whenever } \type(l_\rec)=\inputm  \text{  and  }(l_\rec,\rec)\in P, \cr
\emptyset & \text{otherwise };\end{cases}$ 
\item $\textit{consumed}_{\eme}(P)\bydef \begin{cases}
\{l_\eme\} & \text{whenever }(l_\eme,\eme)\in P,\cr
\emptyset  & \text{otherwise;}\end{cases}$

\item $\textit{created}_{\rec}(P)\bydef\{l\;|\;l\not=l_{\rec}, (l,\rec)\in P\}$ and $\textit{created}_{\eme}(P)\bydef\{l\;|\;l\not=l_{\eme}, (l,\eme)\in P\}$;
\item $\textit{content}_0(P) \bydef \textit{old}(P) \minnum (\xinum(\textit{consumed}_\rec \cup \textit{consumed}_{\eme})) \addnum  (\xinum(\textit{created}_\rec \cup \textit{created}_{\eme}))$;
\item $\textit{content}_1(P) \bydef \textit{update\_trans}(\lrec,\leme)(\textit{content}_0(P))$.
\end{itemize}
\end{minipage}}
\caption{Contents analysis.}
\label{absocc}
\end{figure*}
At the beginning of the concrete computation, 
 each variable $x$ is bound to  the name $(x,\varepsilon)$. 
Besides, the configuration contains one thread at each program point the label of which is in the set $\Extract(\System)$. 
Thus, a thread at program point $l$ is in the computation unit $[b\mapsto (\giveindex(l)(b),\varepsilon)]$.
So, at the beginning of the computation, a computation unit $u$ is either empty, or it contains a thread at each program point $l\in\Extract(\System)$ such that $\piunit(u)=\giveindex(l)$. 
In the abstract, we introduce a primitive $\xinum\in\wp(\var)\rightarrow \abstnum$. For any set $A\in\wp(\var)$, we denote by $\chi(A)$ the characteristic function of $A$ which  maps any variable $v\in \var$ to $1$ whenever $v\in A$, and to $0$ otherwise. We require that $\chi(A)\in \concnum(\xinum(A))$.

\begin{example}[\ccount]
In our abstract domain, the primitive $\xinum$ maps any set $A\subseteq \var$ of variables, to the set of constraints $\{v = 1 \;|\;v\in \var\}\cup\{ v=0 \;|\; v\not\in \var\}$.
\boxexample\end{example}

The abstract state $\initnum$ is defined in Fig.~\ref{initocc} as the function 
mapping any abstract computation unit $a\in\abstractunit$ to the element  $\cupnum\{\xinum(\{x_l\;|\;l\in \Extract(\System),\;\giveindex(l)=a\});\xinum(\emptyset)\}$.

\begin{example}[\ccontentanalysis]
In the shared memory example (e.g.~see Ex.~\ref{memory}), the abstract element $\initnum$ is equal to:
\begin{equation*}
\begin{cases}
\ccreate & \!\!\mapsto \{0\leq x_{\memlaba}\leq 1\}\cup \{v = 0 \;|\;\forall v\in \var\setminus\{x_{\memlaba}\}\},\cr
\cbang_{12} & \!\!\mapsto \left\{\begin{array}{c}\!\!\!
0\leq x_{\memlabl}\leq 1\!\!\!\cr
\!\!\!x_{\memlabl}=x_{\memlabl'}\!\end{array}\right\}\cup \{v = 0 \;|\;\forall v\in \var\setminus\{x_{{\memlabl}};x_{\memlabl'}\}\},\!\!\!\cr
\_ & \!\!\mapsto \{v=0, \forall v\in \var\};\cr
 \end{cases}
\end{equation*}
since we have: $\Extract(\System)=\{1,12,12'\}$, $\giveindex(1)=\ccreate$, and $\giveindex(12)=\giveindex(12')=\cbang_{12}$.
\boxexample\end{example}

\subsection{Transition step}
We consider an abstract element $\tp\in\Numdom$, two program point labels $\lrec$ and $\leme$, and a transition sub-case $(\sim,A)\in\abstractcontext(\lrec,\leme)$. We simulate in the abstract any computation step $\tau$ that matches $C \nonstan{\lambda} C'$, where $\lambda=((\lrec,\leme),(\sim,A))$ (e.g.~see Fig.~\ref{postocc}).

\begin{example}[\ccontentanalysis]
As a running example, we simulate an interaction between a thread at the program point \pp{\memlabe} and 
a thread at the program point \pp{\memlabj}. 
We start from an abstract element $\tp$ such that
the system $\tp([b\mapsto \ccell])$ is made of both the constraints $x_{\memlabb} + x_{\memlabf} + x_{\memlabj} = y_{\memlaba,\memlabm}$ and $0\leq y_{\memlaba,\memlabm} \leq 1$.

We set $\lrec=\memlabe$ and $\leme=\memlabj$.
Thanks to the control flow analysis, we only take into account the transitions where $(\memlabe,\rec)\sim(\memlabf,\rec)$, $(\memlabe,\rec)\sim(\memlabj,\eme)$, $(\memlabe,\rec)\not \sim (\memlabg,\rec)$, $A([(\memlabe,\rec)]_\sim)(b)=\ccreate$, and $A([(\memlabg,\rec)]_\sim)(b)=\addread$. Indeed, results coming from the other cases are ignored thanks to the coalesced product (e.g.~see \ref{product}). 
\boxexample\end{example}

\subsubsection{Is the step possible ?}
First, we check whether the computation step is possible, or not. 
Whenever we have $(\lrec,\rec)\sim (\leme,\eme)$, 
there must be a computation unit $u$ in $C$ such that both  $\piunit(u)=A([(\lrec,\rec)]_\sim)$ and $u$ contains at least one thread at the program point $\lrec$ and one thread at the program point $\leme$; 
whenever we have $(\lrec,\rec)\not\sim (\leme,\eme)$,
there must be two computation units $u_{\rec}$ and $u_{\eme}$ such that: for any $\diamond\in\{\rec;\eme\}$, $\piunit(u_{\diamond})=A([(l_{\diamond},\diamond)]_\sim)$ and  $u_{\diamond}$ contains at least a thread at the program point $l_{\diamond}$.
To check these properties, we require an abstract primitive $\syncnume\in\wp(\var)\rightarrow \abstnum \rightarrow \abstnum$ to check whether some variables may simultaneously take a non-zero value.
For any set $I$ of variables and any abstract element $a\in\abstnum$, 
the set $\{f\in\gammanum(a)\;|\;\forall v\in I,\;f(v)\geq 1\}$ should be included in the concretization $\gammanum(\syncnume(I)(a))$.
If there exists $\diamond\in \{\rec;\eme\}$ such that 
$\syncnume([(l_{\diamond},\diamond)]_{\sim}\cap \{(\lrec,\rec);(\leme,\eme)\})(\tp(A([(l_{\diamond},\diamond)]_\sim)))$ is equal to the bottom element $\botnum$, 
the computation step is not possible, so we define $\abst{\numkey}(\tp,((\lrec,\leme),(\sim,A)))$ as $\botNum$.
Otherwise, we  update the abstraction of any computation unit involved in the computation step.

\begin{example}[\ccontentanalysis]
We know that i) $(\memlabe,\rec)\sim (\memlabj,\eme)$ and ii) $A([(\memlabe,\rec)]_\sim)=[b\mapsto \ccell]$.
We compute $t$ that is defined by the expression   $\syncnume(\{(\memlabe,\rec);(\memlabj,\eme)\})(\tp([b\mapsto \ccell]))$.
The system $t$ is equivalent to the system: 
\begin{equation*}
\begin{cases}
x_{\memlabe}\geq 1,\;x_{\memlabj}\geq 1,\;
x_{\memlabb} + x_{\memlabf} + x_{\memlabj} = y_{\memlaba,\memlabm},\;
0 \leq y_{\memlaba,\memlabm} \leq 1.
\end{cases}
\end{equation*}
 By reduction, we obtain that 
$t$ is equivalent to the system:
\begin{equation*}
\begin{cases}
x_{\memlabe}\geq 1,\;
y_{\memlaba,\memlabm}=x_{\memlabj}=1,\;
x_{\memlabf}=x_{\memlabb}=0.
\end{cases}
\end{equation*}  
This means that the interaction is only enabled when the cell has already been created ($y_{\memlaba,\memlabm}=1$) and when both interacting threads are in the computation unit ($x_{\memlabe}\geq 1$ and $x_{\memlabj}=1$). 
In this case, there is no thread at either the program point \pp{\memlabb}, or at the program point \pp{\memlabf} ($x_{\memlabf}=x_{\memlabb}=0$). 
\end{example}

\subsubsection{Abstracting the former contents of partition classes}

Let us consider a class $P\in\classes$. 
The class $P$ denotes a computation unit $u$ that is transformed during the computation step. 
We first compute an abstraction $\textit{old}(P)$ of the contents of $u$ before the computation step. 
In the case where there exists a pair $(l,\diamond)\in P$ and a key $b\in B$ such that $\giveindex(l)(b)\in \interface(l)\setminus \FN(\cont(l_\diamond))$,  the computation unit maps a key to a fresh name, 
so we can deduce that the computation unit $u$ has been created during the transition step. In such a case, we define $\textit{old}(P)$ as $\xinum(\emptyset)$. Otherwise, we take into account the abstraction of the computation unit and the threads that are required to enable the computation step: we define $\textit{old}(P)$ as $\syncnume(P\cap \{(\lrec,\rec);(\leme,\eme)\})(\tp(A(P)))$.

\begin{example}[\ccontentanalysis]
First, we compute the contents of partition class $[(\memlabe,\rec)]_\sim$ before the computation step.
The  element  $\textit{old}([(\memlabe,\rec)]_\sim)$ is equal to 
$\syncnume([(\memlabe,\rec)]_{\sim}\cap \{(\memlabe,\rec);(\memlabj,\eme)\})(\tp(A([(\memlabe,\rec)]_\sim)))$, so the system $\textit{old}([(\memlabe,\rec)]_\sim)$ contains the constraints  $x_{\memlabe}\geq 1$, $x_{\memlabj}\geq 1$, 
$x_{\memlabb} + x_{\memlabf} + x_{\memlabj} = y_{\memlaba,\memlabm}$, and 
$0 \leq y_{\memlaba,\memlabm} \leq 1$. By reduction, we obtain that the system  $\textit{old}([(\memlabe,\rec)]_\sim)$ is given by the constraints $x_{\memlabe}\geq 1$, $y_{\memlaba,\memlabm}=x_{\memlabj}=1$, $x_{\memlabf}=x_{\memlabb}=0$.  
This means that the interaction is only enable when the cell has already been created ($y_{\memlaba,\memlabm}=1$) and if the interacting threads are in the computation unit ($x_{\memlabe}\geq 1$ and $=x_{\memlabj}=1$). 
In such a case, there is no thread at the program point \pp{\memlabb} or at the program point \pp{\memlabf} ($x_{\memlabf}=x_{\memlabb}=0$). 
\boxexample\end{example}

\begin{example}[\ccontentanalysis]
We now consider a case when  a computation unit is necessarily empty.
We simulate an interaction between a thread at the program point
\pp{\memlaba} and a thread at the program point \pp{\memlabm}.
This way, we set $\lrec=\memlaba$ and $\leme=\memlabm$.
Thanks to the control flow analysis, we only take into account the transitions where $(\lrec,\rec)\sim(\leme,\rec)$  and $A([(\lrec,\rec)]_\sim)(b)=\ccreate$. 
The interaction launches a thread at the program point \pp{\memlabb}.
But, we have  $\giveindex(\memlabb)(b)=\ccell$, $\interface(\memlabb)=\{\ccell;\cnull\}$, and $\FN(\cont(\memlaba))=\{\cnull;\addinit\}$.
So $\ccell\in \interface(\memlabb)\setminus \FN(\cont(\memlaba))$.
Thus we can conclude that $\textit{old}([(\memlabb,\rec)]_\sim)$ is equal to $\xinum(\emptyset)$. This way, the thread is launched in an empty computation unit.
\boxexample\end{example}

\subsubsection{Abstracting the evolution of partition classes}

Then, we compute the set of labels of the threads that are created and consumed in the computation unit $u$.
The input thread is consumed in $u$ only if it is not a resource and if it was in the computation unit $u$: so we define $\textit{consumed}_{\rec}(P)$ as $\{l_\rec\}$ if both  $\type(l_\rec)=\inputm$  and $(l_\rec,\rec)\in P$, and as 
  $\emptyset$ otherwise. The output thread is always consumed (we only check whether it is in $u$, or not): so we define $\textit{consumed}_{\eme}(P)\bydef\{l_\eme\}$ if $(l_\eme,\eme)\in P$, and 
 $\textit{consumed}_{\eme}(P)\bydef\emptyset$ otherwise. The threads that are created during the computation step are dealt with the same way: we define $\textit{created}_{\diamond}(P)\bydef\{l\;|\;l\not=l_{\diamond}, (l,\diamond)\in P\}$, for any $\diamond\in\{\rec;\eme\}$. 

\begin{example}[\ccontentanalysis]
In our running example, the set 
$\textit{consumed}_{\rec}([(\memlabe,\rec)]_\sim)$ is equal to $\{\memlabe\}$, the set $\textit{consumed}_{\eme}([(\memlabe,\rec)]_\sim)$ is equal to $\{\memlabj\}$. Since the constraints $(\memlabe,\rec)\sim (\memlabf,\rec)$ and $(\memlabe,\rec)\not \sim (\memlabg,\rec)$ are satisfied, we can deduce that the set $\textit{created}_{\rec}([(\memlabe,\rec)]_\sim)$ is equal to $\{\memlabf\}$. Last, the set 
  $\textit{created}_{\eme}([(\memlabe,\rec)]_\sim)$ is empty.
\boxexample\end{example}

The abstraction $\textit{content}_0(P)$ of the contents of the computation unit $u$ after the computation step can then be defined as $\textit{old}(P) \minnum (\xinum(\textit{consumed}_\rec \cup \textit{consumed}_{\eme})) \addnum  (\xinum(\textit{created}_\rec \cup \textit{created}_{\eme}))$, where $\addnum$ and $\minnum$ are sound counterparts to the point wise addition and to the point wise subtraction. More precisely, for any $a,b\in\abstnum$ and $\circ\in\{+;-\}$, we have: 
$a\circ^{\sharp} b\in \abstnum$, and the concretization $\gammanum(a\circ^{\sharp} b)$ contains at least all functions $[v\mapsto f(v)\circ g(v)]$ such that: $f\in\gammanum(a)$, $g\in\gammanum(b)$, and for any $v\in\var$, $f(x)\circ g(x)\geq 0$. 

\begin{example}[\ccount]
\hspace*{-1mm}The primitives $+^{\sharp}$ and  $-^{\sharp}$ are both computed pair-wise over the system of affine constraints and over the system of interval constraints. More details can be found in \cite[Chap.~9,\;Sect.~9.3.1]{feret:thesis}..
\boxexample\end{example}

The last step consists in updating the local history of computation units. 
We introduce a primitive $\textit{update\_trans}\in\Labels^2 \rightarrow \abstnum \rightarrow \abstnum$. The function $\textit{update\_trans}(\lambda)$ increments, in the abstract, the value of variable $y_{\lambda}$ and sets the value of variable $z_{\lambda}$ to $1$. So, for any function $f\in\gammanum(a)$, the function $g$ that maps $y_\lambda$ to $f(y_\lambda)+1$, $z_\lambda$ to $1$, and any other variable $v$ to $f(v)$ should be an element of the concretization $\gammanum(\textit{update\_trans}(\lambda)(a))$. Thus, we define $\textit{content}_1(P)$ as $\textit{update\_trans}(\lrec,\leme)(\textit{content}_0(P))$. 

\begin{example}[\ccount]
We can define the primitive $\textit{update\_trans}(\lambda)$ by using the usual transfer functions for assignments in interval constraints (e.g.~see \cite{cc76}) and in affine equalities (e.g.~see \cite{karr}).
\boxexample\end{example}

\begin{example}[\ccontentanalysis]
In our running example, the system 
$\textit{content}_0([(\memlabe,\rec)]_\sim)$ is given by the constraints $x_{\memlabe}\geq 0$, $y_{\memlaba,\memlabm}=1$,  $x_{\memlabj}=0$, 
$x_{\memlabf}=1$, $x_{\memlabb}=0$.  
Then, $\textit{content}_1([(\memlabe,\rec)]_\sim)$ is given by the constraints 
$x_{\memlabe}\geq 0$, $y_{\memlaba,\memlabm}=1$,  $x_{\memlabj}=0$, 
$x_{\memlabf}=1$, $x_{\memlabb}=0$, $y_{\memlabe,\memlabj}\geq 1$, and $z_{\memlabe,\memlabj}=1$.
\boxexample\end{example}

\subsubsection{Updating abstract elements}
We are left to update the abstraction of the computation units whose 
abstract computation unit is $A(P)$. We define, for any $a\in\abstractunit$, $\abst{\numkey}(\tp,((\lrec,\leme),(\sim,A)))(a)$ as $\cupnum \{\tp(a)\}\cup \{\textit{content}_1(P)\;|\;P\in\classes,\;A(P)=a\}$.

\begin{example}[\ccontentanalysis]
We recall the fact that the system $\tp([b\mapsto \ccell])$ entails the affine constraints $x_{\memlabb} + x_{\memlabf} + x_{\memlabj} = y_{\memlaba,\memlabm}$ and the interval constraint $0\leq y_{\memlaba,\memlabm} \leq 1$. 
The class $P=[(\memlabe,\rec)]_\sim$ is the only one such that 
$A(P)=[b\mapsto\ccell]$. Moreover,  the affine constraints $x_{\memlabb} + x_{\memlabf} + x_{\memlabj} = y_{\memlaba,\memlabm}$ and the interval constraint $0\leq y_{\memlaba,\memlabm} \leq 1$ are also entailed by the system $\textit{content}_1(P)$. The analysis discovers that these constraints are invariant.
\boxexample\end{example}

\subsection{Soundness}

Thm.~\ref{contsound} states the soundness of our content analysis.

\begin{theorem}
\label{contsound}
$\tuple{\Numkey}$ is an abstraction.
\end{theorem}

\section{Conclusion}

We have proposed a generic framework for statically inferring properties of mobile systems. This framework is based on thread partitioning: we gather the threads of a mobile system into several classes. The criterion of thread partitioning is left as a parameter. We use the product of an analysis of the dynamic linkage between the threads of a system and an analysis of the number of threads inside each partition class. As a result, we get a polynomial-time (with respect to the length of the initial state) analysis, which succeeds in proving the absence of race conditions in a shared memory  written in the \picalcul.
In \cite[Chap:10]{feret:thesis}, we propose a version of this framework for the \ambient-calculus (see.~Sect.~10.2), and a model independent version (see.~Sect.~10.3). 
We succeed in proving authentication properties in a version \cite{Woo} of the Woo and Lam one-way public-key authentication protocol that is written in the \spicalcul\ \cite{spi}. For that purpose, we partition the threads according to the identities of the principals that have initiated the session. 

Thread partitioning may also be used in \emph{reconfigurable systems} to prove that the system may not switch to a new version until all components have  been installed. For that purpose, we may partition threads according to the version identifier. As future works, we are also  interested in using thread partitioning to refine the type checking of authorization policies \cite{authorization}.

\bibliography{document}
\end{document}